\documentclass[lettersize,journal]{IEEEtran}
\usepackage{amsmath,amsfonts}
\usepackage{algorithmic}
\usepackage{algorithm}
\usepackage{array}
\usepackage{float}
\usepackage{textcomp}
\usepackage{url}
\usepackage{verbatim}
\usepackage{graphicx}
\usepackage{float}
\usepackage{subfigure}
\usepackage{cite}
\usepackage{verbatim}
\usepackage{siunitx}
\usepackage{caption}
\usepackage{booktabs} 
\usepackage{multirow}
\usepackage{mathrsfs}
\usepackage{epstopdf}
\usepackage{epsfig}
\usepackage{makecell}
\usepackage{amsmath}

\usepackage[numbers]{natbib}
\begin{document}

\title{Energy Efficiency Optimization Method of WDM Visible Light Communication System for Indoor Broadcasting Networks}

\author{Dayu Shi,~\IEEEmembership{Member,~IEEE,} $^{*}$Xun Zhang,~\IEEEmembership{Senior Member,~IEEE,} Ziqi Liu, Xuanbang Chen,\\ Jianghao Li, Xiaodong Liu,~\IEEEmembership{Member,~IEEE,},  and William Shieh,~\IEEEmembership{Fellow,~IEEE,}

\thanks{D. Shi, and W. Shieh are with the School of Engineering, Westlake University, Hangzhou, 310030, China (e-mail: shidayu@westlake.edu.cn, shiehw@westlake.edu.cn).}
\thanks{ X. Zhang, Z. Liu, and X. Chen are with the Institut supérieur d'électronique de Paris, Paris 75006, France (e-mail: xun.zhang@isep.fr, ziqi.liu@isep.fr, xuanbang.chen@ext.isep.fr).}
\thanks{X. Liu are with the School of Information Engineering, Nanchang University, Nanchang 330031, China (e-mail: xiaodongliu@ncu.edu.cn).}
\thanks{J. Li is with the School of Physics and Electronics, Shandong Normal University, Jinan 250014, China (e-mail: sdnuljhao@sdnu.edu.cn).}
}

\markboth{Journal of \LaTeX\ Class Files,~Vol.~14, No.~8, August~2021}%
{Shell \MakeLowercase{\textit{et al.}}: A Sample Article Using IEEEtran.cls for IEEE Journals}


\maketitle

\begin{abstract}

This paper introduces a novel approach to optimize energy efficiency in wavelength division multiplexing (WDM) Visible Light Communication (VLC) systems designed for indoor broadcasting networks. A physics-based LED model is integrated into system energy efficiency optimization, enabling quantitative analysis of the critical issue of VLC energy efficiency: the nonlinear interplay between illumination and communication performance. The optimization jointly incorporates constraints on communication quality of each channel, and illumination performance, standardized by the International Commission on Illumination (CIE). The formulated nonlinear optimization problem is solved by the Sequential Quadratic Programming (SQP) algorithm in an experiment-based simulation. An integrated Red-Green-Blue-Yellow Light Emitting Diode (RGBY-LED) is measured for model calibration and three different scenarios are simulated to evaluate the generality of the proposed method. Results demonstrate a double enhancement in performance and a high versatility in accommodating various scenarios. Furthermore, it highlights the importance of balancing communication and illumination imperatives in VLC systems, challenging conventional perceptions focused solely on minimizing power consumption.

\end{abstract}


\section{Introduction}
\IEEEPARstart{T}{th} white paper on the sixth generation (6G) architecture landscape from the European perspective is delivered recently \cite{whitepaper}. It proposes the main novel trends and principles that will form the backbone of future 6G network architecture and discusses along different axes related to 6G features. One of the most significant architectural enablers is the sustainable network. According to the Sustainable Development Goals (SDG) of the United Nations(UN), the white paper proposes a threefold sustainability from the societal, economic, and environmental targets, which requires reducing more than 30\% $CO_{2}$ emissions in 6G powered sectors of society,  more than 30\% total cost of ownership, and more than 90\% energy transmitted per bit, respectively. 

With the strict energy efficiency requirements of the 6G network, revolutionized Internet of Things (IoT) applications require ten or even a hundred times the performance improvement compared to the the fifth generation (5G) era \cite{nguyen20216g}. The expected Key Performance Indicators (KPIs) include the dense of connection $10^{7}$ devices/$km^{2}$, the mobile traffic capability 1 $Gbits/s/m^{2}$, the network latency 10-100 $\mu s$, the energy efficiency 1 $Tb/J$, etc \cite{6G_kpi,6G_kpi2}.
  
Meeting the increasing demands of IoT applications while balancing energy efficiency through conventional wireless communication technologies is a formidable challenge \cite{chowdhury20206g}. Visible Light Communication (VLC), modulating visible light to transmit information and simultaneously provide illumination, offers a promising solution \cite{matheus2019visible}. Numerous studies have highlighted the benefits of VLC systems \cite{chi2020visible,ariyanti2020visible,shi2022ai}. From a sustainability standpoint, the most compelling feature of VLC systems is energy efficiency, which leverages existing illumination resources for indoor communication services. To maximize this superiority, a complete definition and accurate analysis of VLC energy efficiency are expected for massive-IoT networks in the 6G era.

Existing research on the energy efficiency analysis and optimization of VLC systems are considered from three perspectives. First, some researchers focuses on hybrid systems, encompassing VLC combined with Radio Frequency (RF) \cite{c3:RF/VLC1,c3:RF/VLC2}, Power Line Communication (PLC) \cite{c3:vlc/plc/rf}, and Long Term Evolution (LTE) \cite{c3:vlc/lte}. Within these contexts, researchers aim to integrate VLC systems with other technologies while maintaining optimal global energy efficiency. 

Subsequently, researchers address the energy efficiency of VLC systems within specific scenarios such as Non-Orthogonal Multiple Access (NOMA) \cite{c3:noma}, heterogeneous networks \cite{c3:hetero}, Multi-Input Multi-Output (MIMO), and broadcasting \cite{c3:down}. These studies elucidate the challenges in enhancing VLC energy efficiency in multi-user environments and offer optimized solutions for resource allocation, including power, subchannels, and spectrum.

\begin{table*}[!htbp]
    \centering
   \renewcommand\arraystretch{1.4}
    \caption{ Representative Existing Research of VLC System Energy Efficiency}

    \begin{tabular}{|c|c|c|c|c|c|}
    \hline
        \textbf{Paper} & \textbf{Objective System} & \textbf{Objective Scenario} & \textbf{Objective Variable} & \textbf{LED Model} & Illumination Requirements\\
        \hline
         \cite{c3:soa/rfvlc} & Hybrid VLC/RF & MIMO & Signal Power & Constant & None \\
         \hline
         \cite{c3:vlc/plc/rf} & Hybrid VLC/RF/PLC & Broadcasting & Signal Power & Constant & None\\
         \hline
         \cite{c3:ledarrary} & Multi-LED VLC & Broadcasting & Signal Power & Constant & Illumination Intensity\\
         \hline
         \cite{c3:vlcillust} & VLC & MIMO & Signal Power & Constant & Illumination Intensity\\
         \hline
         \cite{c3:mashuai} & VLC & MIMO & Amplifier Gain & Linear & Illumination Intensity\\
         \hline
         \cite{c3:deng} & VLC &MIMO & Signal Power & Low-Pass & None\\
         \hline
         \cite{c3:diming} & VLC & Broadcasting & LED's Current & Empirical Formula & Illumination Intensity\\
         \hline
         \cite{c3:tang} & VLC/SLIPT & MIMO & LED's Current/Signal Power& Linaer & Illumination Intensity\\
         \hline
    \end{tabular}

    \label{c3:t1}
    
\end{table*}
Additionally, the other research refines the specific optimization variables, most notably input signal power \cite{c3:power}, LED Direct Current (DC) \cite{c3:current}, and amplifier gain at the transmitter \cite{c3:amp}. These variables are critical for achieving the best energy-efficient configuration of VLC systems. Concurrently, as VLC serves dual functions, illumination and communication, both requirements are integrated as constraints in the optimization process \cite{c3:limits}. For a comprehensive understanding of the current research of VLC system energy efficiency, the representative existing research is detailed in Tab. \ref{c3:t1}.

The related works exhibit excellent performance in optimizing VLC system performance, whereas there still remain several research gaps toward applying to practical VLC system energy efficiency optimization.

First, existing research predominantly treats VLC systems as electronic systems, thereby focusing on optimizing the signal power to achieve optimal energy efficiency. Whereas VLC systems work as electro-optical systems, providing both illumination and communication services simultaneously. Its energy efficiency optimization must concurrently evaluate the illumination and communication power consumption to solve the optimal configuration of the system, reaching the global maximum energy efficiency. 

Second, empirical measurements confirm a nonlinear interplay between illumination and communication performance in VLC systems, attributable to the LED's intrinsic nonlinearity \cite{c3:shidayu,c3:interplay}. Current research overlooks this interplay due to their LED models established by a constant parameter or any mathematical formula without physical significance. Few researchers employ a physics-based LED in their system model to quantify this interplay in their optimization, resulting in suboptimal optimization results. 

Third, the majority of existing methods have not adequately incorporated indoor illumination requirements, such as illuminance and color, into their optimization algorithms. These values are regulated by the International Commission on Illumination (CIE) standards to ensure human ocular safety across various scenarios \cite{smith1931cie}. Thus, the indoor illumination indicators should be formulated into functions and applied to the optimization as constraints.

Last, the polychromatic LED-based wavelength division multiplexing (WDM) VLC systems perform extremely high data rates and illumination efficiency \cite{c3:wdmrgby}. Especially the Red-Green-Blue-Yellow (RGBY)-LED-based WDM VLC transmission reached the data rate of 15.73 $Gbit/s$ \cite{c3:wdmref1}, becoming the most potential architecture in future VLC system applications. Additional merits include improved signal quality, facilitated by optimizing multiple color channels for varying transmission conditions, and increased fault tolerance through redundant data pathways \cite{c3:wdmref3}. However, the existing optimization method fails to be implemented for WDM VLC systems due to the lack of an appropriate system model.

This paper proposed a novel optimization method for the RGBY-LED-based WDM VLC broadcasting system. Addressing the limitations identified in existing research, the contributions of the proposed optimization method are summarized as follows.

    \begin{itemize} 
     \item  A physics-based LED model, developed in our previous work \cite{shi2023physics}, is employed to accurately characterize the nonlinear interplay between illumination and communication performance of VLC systems. The LED model integrated with the channel and receiver model forms a WDM VLC system model, serving as the fundamental function of energy efficiency optimization. \\
     
    \item Based on the system model, the proposed optimization method jointly evaluates the power consumption of illumination and communication services. The illuminance of each LED, the Bit Error Rate (BER) of each color channel, and the visual color of the mixed light are formulated as constraint functions applied to the optimization problem, forcing the results satisfying both indoor illumination standards and communication quality requirements across diverse scenarios.  \\

    \item The formulated nonlinear optimization problem is solved by the Sequential Quadratic Programming (SQP) algorithm in an experiment-based simulation. An integrated RGBY-LED is measured in a VLC transmission testbed to calibrate the LED model applied to the optimization. An indoor VLC broadcasting scenario with different use-cases are simulated to evaluate the performance and generality of the proposed method.

\end{itemize}{}

\section{WDM VLC System Model}



\begin{figure*}[htbp!]
\centering
\includegraphics[scale = 0.6]{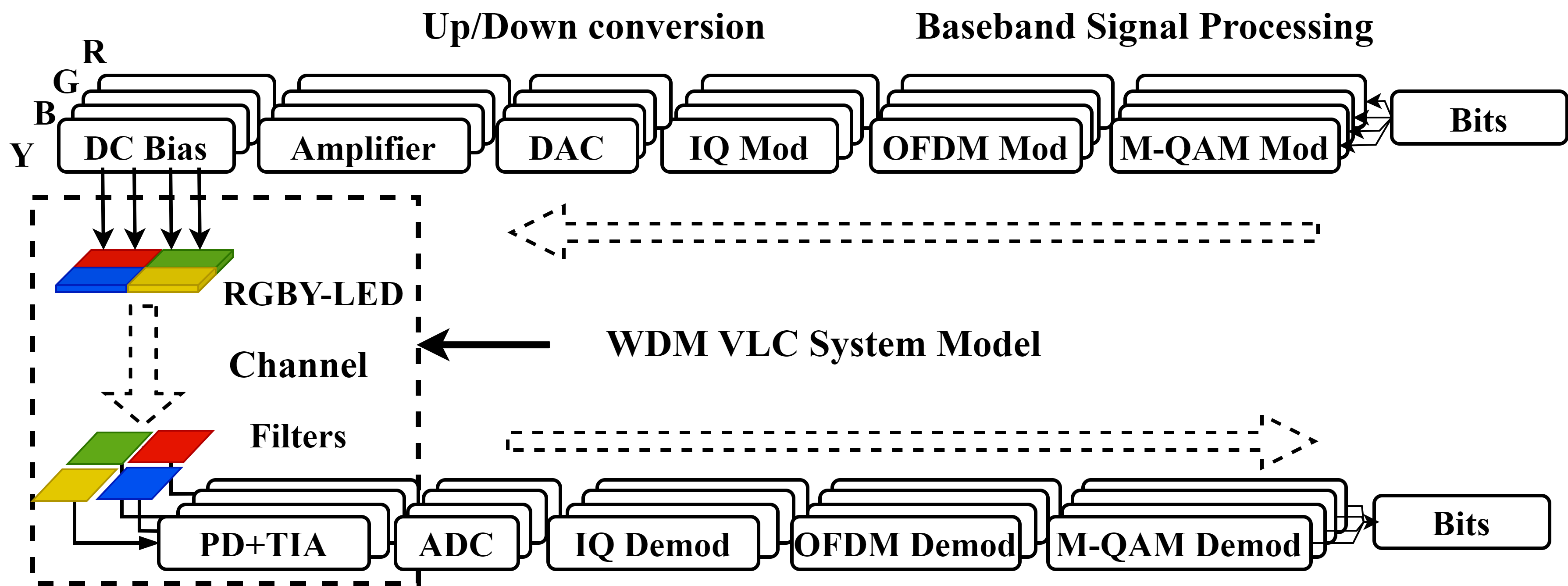}
\caption{Schematic diagram of the RGBY-LED-based WDM VLC broadcasting system. }
\label{c3:wdm system model}
\end{figure*}

Fig. \ref{c3:wdm system model} delineates the architecture of the RGBY-LED-based WDM VLC broadcasting system. This system incorporates four sub-LEDs—red, green, blue, and yellow—into a single RGBY-LED unit. Data are concurrently transmitted across these four color channels, forming a WDM mechanism. Additionally, the sub-LEDs individually generate calibrated ratios of red, green, blue, and yellow light, which combine to create white light suitable for indoor illumination.

In the transmission process, the bitstream is partitioned into four separate flows. Each flow is mapped onto M-order Quadrature Amplitude Modulation (QAM) and Orthogonal Frequency Division Multiplexing (OFDM) symbols to create the baseband signals. Utilizing In-phase and Quadrature (IQ) modulation and applying a Digital Analog Converter (DAC) to each channel, these baseband signals are upconverted to higher frequencies and transformed into analog signals. Within each channel, the signal is amplified and added with DC components before transmitting by the corresponding sub-LED. Over the optical wireless channel, the signal of each color is filtered and then detected by the Photodiode (PD), which converts the optical signals back into current signals. A Trans-Impedance Amplifier (TIA) is used to amplify these current signals and convert them into voltage signals. After undergoing down-conversion and demodulation, the original bits are reconstructed.

In this paper, the established WDM VLC system model consists of the RGBY-LED model, channel model, and receiver model,  covering the main configurable components of VLC systems. Without loss of generality, the rest are temporarily considered constantly and ideally. The detailed derivation of the system model is separately introduced as follows.

\begin{figure}[htbp!]
\centering
\includegraphics[scale=0.54]{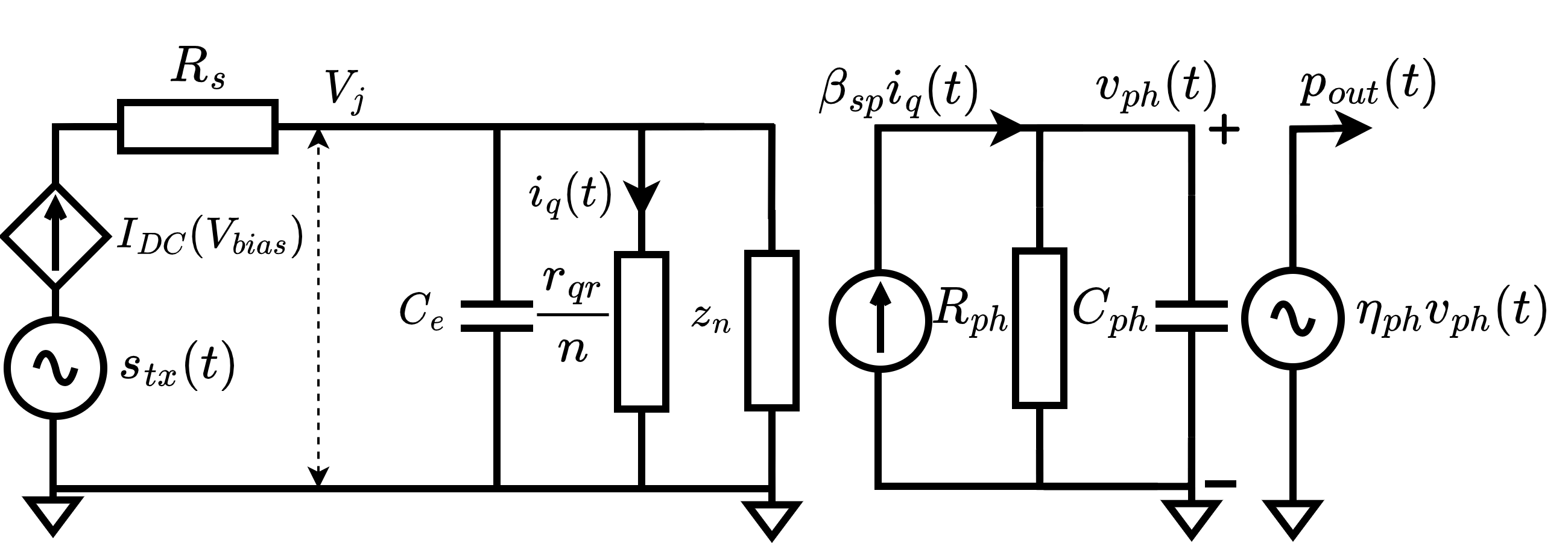}
\caption{Simplified small signal equivalent circuit of LEDs.}
\label{c3:LED}
\end{figure}

\subsection{GaN MQW LED Model} \label{sec:A_1}


A monochromatic LED model is developed and demonstratively verified in our previous work \cite{shi2023physics}. Based on the theory, the model of RGBY-LED is topologized in this section. Considering each quantum well has the same structure and size, the small signal equivalent circuit of the LED is simplified as Fig. \ref{c3:LED}. For illustrative convenience, assuming a monochromatic LED featured $n$ quantum wells structure, at bias voltage $V_{bias}$, its output illuminance ($\Phi_{LED}$) and signal response ($h_{led}$) are represented by:

\begin{equation}
\begin{aligned}
  \Phi_{LED} (V_{bias})=\frac{638}{A_{rec}}\int_{380}^{830}\mathcal{V}(\lambda) P_{L}(V_{bias},\lambda) d\lambda
  \end{aligned}
\label{c3:lumin}
\end{equation}

\begin{equation}
\begin{aligned}
&h_{led}(V_{bias},t)=\mathcal{L}^{-1}(H(V_{bias},\boldsymbol{s}))\cdot\exp \bigg(\frac{-4(\lambda-\lambda_{0})^{2}}{(\Delta \lambda)^{2}}\bigg)\\
&=\frac{a_{1}(V_{j}) }{a_{4}(V_{j})-a_{2}\cdot a_{3}(V_{j})}\cdot \Bigg[\exp\bigg(\frac{-a_{3}(V_{j})\cdot t}{a_{4}(V_{j})}\bigg)-\exp\bigg(\frac{-t}{a_{2}}\bigg)\Bigg] \\
&\cdot \exp \bigg(\frac{-4(\lambda-\lambda_{0})^{2}}{(\Delta \lambda)^{2}}\bigg)
\label{c3:hled}
\end{aligned}
\end{equation}

 \begin{equation}
\begin{aligned}
P_{L}(V_{bias},\lambda)= a_{5}(V_{j}) \cdot \exp\bigg(\frac{-4(\lambda-\lambda_{0})^{2}}{(\Delta \lambda)^{2}}\bigg)
\label{c3:Pout_illumi}
\end{aligned}
\end{equation}

\begin{equation}
\begin{aligned}
&H_{led}(V_{bias},\boldsymbol{s})=\frac{p_{out}(\boldsymbol{s})}{s_{tx}(\boldsymbol{s})}=\frac{a_{1}}{(1+a_{2}\cdot \boldsymbol{s})(a_{3}+a_{4}\cdot \boldsymbol{s})}
\label{c3:Hled}
\end{aligned}
\end{equation}

\begin{equation}
V_{j}=V_{bias}-R_{s}I_{DC}
\label{Vj}
 \end{equation}

\begin{equation}
I_{DC}=I_{s}(\exp(\frac{q(V_{bias}-R_{s}I_{DC})}{\eta k_{B}T})-1)
\label{c3:shockley}
 \end{equation}
where, $P_{L}$ represents the LED's optical spectrum, calculated as Eq. \eqref{c3:Pout_illumi}, and $\mathcal{V}$ is the luminosity function defined by the CIE \cite{c3:cie}. The power transfer function of the LED is denoted as $H_{led}$, derived as Eq. \eqref{c3:Hled}, and the inverse Laplace Transform operator is $\mathcal{L}^{-1}$. Parameters $\lambda$, $\lambda_{0}$, and $\Delta\lambda$ refer to the wavelength range, peak wavelength, and spectrum bandwidth of the emitted light. The power distribution of the LED's emitted light is modeled using a Gaussian distribution \cite{c3:optspect}.  To clearly illustrate, the equations describe the LED's internal physics are replaced by the parameter functions $a_{1}$ to $a_{5}$, and their expansion are detailed in the Appendix.

For an RGBY-LED, the output illuminance ($\Phi_{LED}^{R,G,B,Y}$), in units of Lux (lx), and the four-channel signal response ($h_{led}^{R,G,B,Y}$) are calculated by: 

\begin{equation}
\begin{aligned}
  \Phi_{LED}^{R,G,B,Y}=\frac{638}{A_{rec}}\int_{380}^{830}\mathcal{V}(\lambda) P_{L}^{R,G,B,Y}(V_{bias}^{R,G,B,Y}) d\lambda
  \end{aligned}
  \label{c3:Phi rgby}
\end{equation}

\begin{equation}
\begin{aligned}
h_{led}^{R,G,B,Y}(V_{bias}^{R,G,B,Y},t)=\left(                 
  \begin{array}{c}   
    h_{led}^{R}(V_{bias}^{R},t)\\  
    h_{led}^{G}(V_{bias}^{G},t)\\
    h_{led}^{B}(V_{bias}^{B},t)\\ 
    h_{led}^{Y}(V_{bias}^{Y},t)\\ 
  \end{array}
\right)
\label{c3:ht RGBY}
\end{aligned}
\end{equation}

To further characterize the illumination performance of the RGBY-LED, the visual color of the mixed light is evaluated by the Correlated Color Temperature (CCT). An empirical function \cite{c3:CCTref} is used to calculate its value as:

\begin{equation}
\begin{aligned}
  CCT=437\cdot \hat{n}^3 + 3601\cdot \hat{n}^2 + 6861\cdot \hat{n} + 5517
  \end{aligned}
  \label{c3:CCT}
\end{equation}

\begin{equation}
\begin{aligned}
  \hat{n}=\frac{\hat{x}-0.3320}{0.1858-\hat{y}}
  \end{aligned}
  \label{c3:lumin2}
\end{equation}

\begin{equation}
\begin{aligned}
  \hat{x}=\frac{\int_{380}^{830}\bar{x}(\lambda)P_{L}^{R,G,B,Y}(\lambda)d\lambda}{\int_{380}^{830}P_{L}^{R,G,B,Y}(\lambda)(\bar{x}(\lambda)+\bar{y}(\lambda)+\bar{z}(\lambda))d\lambda}
  \end{aligned}
  \label{c3:CCT x}
\end{equation}

\begin{equation}
\begin{aligned}
  \hat{y}=\frac{\int_{380}^{830}\bar{y}(\lambda)P_{L}^{R,G,B,Y}(\lambda)d\lambda}{\int_{380}^{830}P_{L}^{R,G,B,Y}(\lambda)(\bar{x}(\lambda)+\bar{y}(\lambda)+\bar{z}(\lambda))d\lambda}
  \end{aligned}
  \label{c3:CCT y}
\end{equation}

 \begin{equation}
\begin{aligned}
V_{bias}^{R,G,B,Y}=(V_{bias}^{R},V_{bias}^{G},V_{bias}^{B},V_{bias}^{Y})
\label{c3:vbiasrgby}
\end{aligned}
\end{equation}

 \begin{equation}
\begin{aligned}
P_{L}^{R,G,B,Y}(V_{bias}^{R,G,B,Y})&= P_{L}^{R}(V_{bias}^{R},\lambda)+P_{L}^{G}(V_{bias}^{G},\lambda)\\
&+P_{L}^{B}(V_{bias}^{B},\lambda)+P_{L}^{Y}(V_{bias}^{Y},\lambda)
\label{c3:PL RGBY}
\end{aligned}
\end{equation}

Where, $V_{bias}^{R,G,B,Y}$ represents the bias voltage applied to each color LED as shown in Eq. \eqref{c3:vbiasrgby}. $P_{L}^{R,G,B,Y}(V_{bias}^{R,G,B,Y})$ denotes the total output power of the RGBY-LED, in units of Watt. $x(\lambda)$, $y(\lambda)$, $z(\lambda)$ are the color matching functions provided by CIE \cite{c3:cie}.

\begin{figure}[htbp!]
\centering
\includegraphics[scale=0.6]{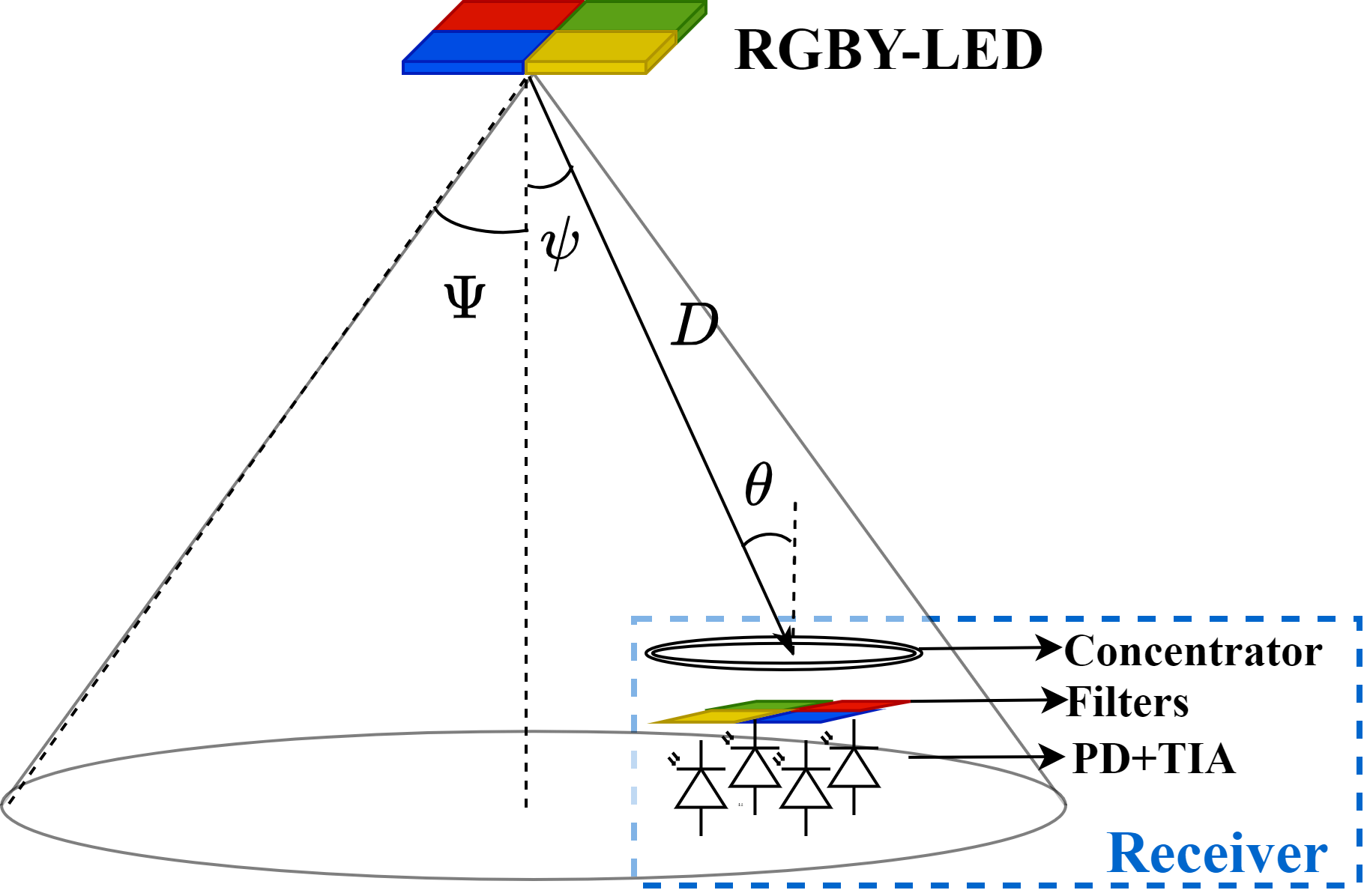}
\caption{Schematics of the channel and receiver.}
\label{c3:fig:channel}
\end{figure}

\subsection{Channel and Receiver Model}
The schematic diagram of the channel and receiver in the WDM VLC system is presented in Fig. \ref{c3:fig:channel}. In the channel model, the optical signal is assumed to propagate in a LOS manner, and a Lambertian radiation model \cite{c3:channel} characterizes the path loss during light propagation, given by:

\begin{equation}
 h_{c}=\bigg\{
\begin{array}{ll}
\delta(t)\cdot \frac{(\mu+1)A_{rec}\cdot cos(\psi)^{\mu}cos(\theta)}{2\pi D^2} &  {0\leq \psi \leq \psi_{FOV}}\\
0 &  \text{otherwise} 
\end{array} 
\label{c3:Lamber}
\end{equation}

\begin{equation}
    \mu=\frac{\ln2}{\ln(\cos{\Psi})}
    \label{c3:mu}
\end{equation}

Parameters $\psi$ and $\theta$ represent the transmit and receive angles, while $D$, $\mu$, and $A_{rec}$ refer to the distance between the transmitter and receiver, Lambertian radiant order, and the effective receiver area, respectively. Symbols $\Psi$ and $\psi_{FOV}$ denote the half-power angle of the LED and the FOV (Field of View) of the receiver.

The receiver setup includes an optical concentrator, an optical filter, and four PDs integrated with TIAs for each channel. In the receiver model, the received signal is computed by integrating the photo-sensitivity of the PDs ($\kappa(\lambda)$) with the filtered optical signal, then multiplying by the gain of the TIA ($G_{TIA}$) and optical concentrator ($G_{opt}$). The optical filters for each channel are represented as $H_{filter}$, and an ideal rectangular filter is used in the model. Parameters $\Gamma_{p}$, $\hat{\lambda}_{c}$, and $\Delta \hat{\lambda}$ denote the filter loss, peak wavelength, and bandwidth, respectively.

\begin{equation}
\begin{aligned}
h_{rx}^{R,G,B,Y}(\lambda)&= \left(                 
  \begin{array}{c}   
      h_{rx}^{R}(\lambda)\\  
    h_{rx}^{G}(\lambda)\\
      h_{rx}^{B}(\lambda)\\ 
     h_{rx}^{Y}(\lambda)\\ 
  \end{array}
\right) 
    \label{c3:Rx}
\end{aligned}
\end{equation}

\begin{equation}
h_{pd}(\lambda)= G_{TIA} G_{opt}\cdot \int \kappa(\lambda)\cdot H_{filter}(\lambda) d\lambda
\end{equation}

\begin{equation}
H_{filter}(\lambda)=\Gamma_{p}\mathrm{rect}(\frac{\lambda-\hat{\lambda}_{c}}{\Delta \hat{\lambda}})
\end{equation}

\begin{equation}
\mathrm{rect}(\lambda)=\left\{
\begin{aligned}
1 &  & \left| \lambda \right|	\leq  \frac{1}{2}\\
0 &  & \left| \lambda\right| >  \frac{1}{2}  \\
\end{aligned}
\right.
\end{equation}

Combining the above RGBY-LED model, channel model, and receiver model, the illuminance at the receiver side ($\Phi_{Rx}^{TL}$) and the response of WDM VLC system ($h_{vlc}^{R,G,B,Y}$) are represented by Eq. \eqref{c3:Phi rx} and \eqref{c3:hvlc}, respectively. While the operators $\times$ and $\otimes$ denote the matrix multiplication and convolution.

\begin{equation}
\Phi_{Rx}^{TL}=\int_{-\infty}^{\infty}\Phi_{Rx}^{R,G,B,Y}\cdot h_{c}(t,D, \psi,\theta) dt
\label{c3:Phi rx}
\end{equation}

\begin{equation}
\begin{aligned}
h_{vlc}^{R,G,B,Y}&=\left(                 
  \begin{array}{c}   
     h_{vlc}^{R}(V_{bias}^{R},t,\psi,\theta,D,\lambda)\\
   h_{vlc}^{G}(V_{bias}^{G},t,\psi,\theta,D,\lambda)\\ 
    h_{vlc}^{B}(V_{bias}^{B},t,\psi,\theta,D,\lambda)\\
    h_{vlc}^{Y}(V_{bias}^{Y},t,\psi,\theta,D,\lambda)
  \end{array}
\right)\\
&=\left(                 
  \begin{array}{cccc}   
     h_{rx}^{R} &h_{rx}^{R} &h_{rx}^{R}& h_{rx}^{R}\\  
     h_{rx}^{G}&h_{rx}^{G}&h_{rx}^{G}&h_{rx}^{G}\\
     h_{rx}^{B}&h_{rx}^{B}&h_{rx}^{B}&h_{rx}^{B}\\ 
     h_{rx}^{Y}&h_{rx}^{Y}&h_{rx}^{Y}&h_{rx}^{Y}\\
  \end{array}
\right) \times  \left(                 
  \begin{array}{c}   
    h_{led}^{R} \otimes h_{c}\\  
    h_{led}^{G}\otimes h_{c}\\
    h_{led}^{B}\otimes h_{c}\\ 
    h_{led}^{Y}\otimes h_{c}\\ 
  \end{array}
\right) 
\label{c3:hvlc}
\end{aligned}
\end{equation}

\section{WDM VLC System Energy Efficiency}

As a combined illumination and communication system, the WDM VLC system exhibits unique properties that differentiate its energy efficiency calculation from conventional wireless communication systems. In this context, the denominator of the efficiency expression should encompass not only the power expended during data transmission but also factor in the illumination power. Furthermore, the WDM system accommodates multiple channels that concurrently transmit data. Given the distinct characteristics of each color LED, individual channel configurations are adjusted to optimize performance \cite{c3:wdmrgby}. As a result, the channel capacity of each color is computed individually and subsequently aggregated to yield the overall system throughput. Each channel is simultaneously optimized to ensure global performance, adhering to constraints imposed by both communication Quality of Service (QoS) and indoor illumination requirements.

\subsection{Energy Efficiency Definition}
The energy efficiency of the WDM VLC system ($EE^{TL}$) is defined as the total channel capacity ($C^{TL}$) of the R, G, B, and Y channels divided by the total power consumption for illumination ($P_{ill}^{R,G,B,Y}$) and communication ($P_{com}^{R,G,B,Y}$), given by:

\begin{equation}
\begin{aligned}
EE^{TL}&=\frac{C^{TL}}{P_{com}^{R,G,B,Y}+P_{ill}^{R,G,B,Y}}\\
&=\frac{C^{R}+C^{G}+C^{B}+C^{Y}}{P_{com}^{R,G,B,Y}+P_{ill}^{R,G,B,Y}}
\label{C3:EE TL}
\end{aligned}
\end{equation}

\begin{equation}
\begin{aligned}
  C^{TL}&=B^{R}\cdot\log_{2}(1+\frac{(s_{rx}^{R}(t))^{2}}{\sigma_{th}^{2}+\sigma_{shot}^{2}})\\
  &+B^{G}\cdot\log_{2}(1+\frac{(s_{rx}^{G}(t))^{2}}{\sigma_{th}^{2}+\sigma_{shot}^{2}})\\
  &+B^{B}\cdot\log_{2}(1+\frac{(s_{rx}^{B}(t))^{2}}{\sigma_{th}^{2}+\sigma_{shot}^{2}})\\
  &+B^{Y}\cdot\log_{2}(1+\frac{(s_{rx}^{Y}(t))^{2}}{\sigma_{th}^{2}+\sigma_{shot}^{2}})
  \label{c3:shannon}
  \end{aligned}
\end{equation}

Considering the thermal noise ($\sigma_{th}$) and shot noise  ($\sigma_{shot}$) at the receiver \cite{c3:noise}, the channel capacity $C$ is expressed by Shannon's law \cite{c3:sref} using the Signal Noise Ratio (SNR) in Eq. \eqref{c3:shannon}.

\begin{equation}
\begin{aligned}
s_{rx}^{R,G,B,Y}&=\left(                 
  \begin{array}{c}   
    s_{rx}^{R}\\  
    s_{rx}^{G}\\
    s_{rx}^{B}\\ 
    s_{rx}^{Y}\\ 
  \end{array}
\right)=\left(                 
  \begin{array}{c}   
    h_{vlc}^{R}\otimes s_{tx}^{R}\cdot G_{amp}^{R}\\  
    h_{vlc}^{G}\otimes s_{tx}^{G}\cdot G_{amp}^{G}\\
    h_{vlc}^{B}\otimes s_{tx}^{B}\cdot G_{amp}^{B}\\ 
    h_{vlc}^{Y}\otimes s_{tx}^{Y}\cdot G_{amp}^{Y}\\ 
  \end{array}
\right)
\end{aligned}
\label{c3:rxrgby}
\end{equation}

\begin{equation}
\begin{aligned}
P_{ill}^{R,G,B,Y}&=V_{bias}^{R}\cdot I_{DC}^{R}+V_{bias}^{G}\cdot I_{DC}^{G}\\
&+V_{bias}^{B}\cdot I_{DC}^{R}+V_{bias}^{Y}\cdot I_{DC}^{Y}
\label{Pill}
  \end{aligned}
\end{equation}

\begin{equation}
\begin{aligned}
P_{com}^{R,G,B,Y}&=(s_{tx}^{R}G_{amp}^{R})^{2}+(s_{tx}^{G}G_{amp}^{G})^{2}\\
&+(s_{tx}^{B}G_{amp}^{B})^{2}+(s_{tx}^{Y}G_{amp}^{Y})^{2}
\label{Pcom}
  \end{aligned}
\end{equation}

Assuming the generated signals after DAC processing are represented by $s_{tx}$. The following amplifiers increase its power in $G_{amp}$ times. The received signal $s_{rx}^{R,G,B,Y}$ is formulated by Eq. \eqref{c3:rxrgby} and the power consumption of the illumination and communication of four LEDs are calculated as Eq. \eqref{Pill} and \eqref{Pcom}.

\subsection{Energy Efficiency Optimization}

In the context of the energy efficiency definition for the WDM VLC system, the bias voltages ($V_{bias}^{R,G,B,Y}$) directly influence signal response and illumination intensity. Meanwhile, the amplifier gains ($G_{amp}^{R,G,B,Y}$) controls the initial signal power of the VLC system, significantly impacting the magnitude of the SNR. As these parameters operating at the system's transmitter, they can be dynamically reconfigured based on the user's requirements. Therefore, they are selected as optimization variables, given by:

\begin{equation}
    \boldsymbol{V_{bias}^{R,G,B,Y}}=[\boldsymbol{V_{bias}^{R}},\boldsymbol{V_{bias}^{G}},\boldsymbol{V_{bias}^{B}},\boldsymbol{V_{bias}^{Y}}]
    \label{c3:opti var 1}
\end{equation}

\begin{equation}
    \boldsymbol{G_{amp}^{R,G,B,Y}}=[\boldsymbol{G_{amp}^{R}},\boldsymbol{G_{amp}^{G}},\boldsymbol{G_{amp}^{B}},\boldsymbol{G_{amp}^{Y}}]
    \label{c3:opti var 2}
\end{equation}

\begin{equation}
\begin{aligned}
&\mathbb{EE^{TL}}( \boldsymbol{V_{bias}^{R,G,B,Y}},\boldsymbol{G_{amp}^{R,G,B,Y}})=\\
&\frac{C^{R,G,B,Y}(\boldsymbol{V_{bias}^{R,G,B,Y}},\boldsymbol{G_{amp}^{R,G,B,Y}})}{P_{ill}^{R,G,B,Y}(\boldsymbol{V_{bias}^{R,G,B,Y}})+P_{com}^{R,G,B,Y}(\boldsymbol{G_{amp}^{R,G,B,Y}})}
\label{c3:opt obj}
\end{aligned}
\end{equation}

The objective function of the energy efficiency optimization is represented by Eq. \eqref{c3:opt obj}.  In order to guarantee both the QoS of data transmission and compliance with indoor illumination standards during optimization, several constraints are outlined as follows:
\begin{itemize}
 
     \item Adequate Illuminance: The illuminance at the user's location ($\Phi_{Rx}^{TL}$) must exceed the user-required illuminance ($\Phi_{Rx}^{req}$).
     \\
    \item Controlled Light Color: Excessive levels of blue light can be detrimental to human eyes \cite{c3:eye}. The CCT of the RGBY-LED should remain below the upper limit set by indoor illumination standards \cite{c3:cie}.
     \\
    \item Assured Data Transmission QoS: The BER of each channel should remain below the thresholds specified by the communication protocol requirements.

\end{itemize}{}

\subsection{Problem Formulation} \label{sec:A_2}
Based on the previously stated definitions, the objective function, and the constraints involved in energy efficiency optimization for the WDM VLC system, the optimization problem incorporating these constraints can be formulated as:

\begin{equation}
\begin{aligned} 
&\max \quad \mathbb{EE^{TL}}( \boldsymbol{V_{bias}^{R,G,B,Y} },\boldsymbol{G_{amp}^{R,G,B,Y}})\\
&\begin{array}{r@{\quad}r@{}l@{\quad}l}
s.t. &\Phi_{Rx}( \boldsymbol{V_{bias}^{R,G,B,Y}})&\geq \Phi_{Rx}^{req} \\
     &BER^{R}(\boldsymbol{V_{bias}^{R}}, \boldsymbol{G_{amp}^{R}})&\leq BER^{req}\\
     &BER^{G}(\boldsymbol{V_{bias}^{G}}, \boldsymbol{G_{amp}^{G}})&\leq  BER^{req}\\
     &BER^{B}(\boldsymbol{V_{bias}^{B}}, \boldsymbol{G_{amp}^{B}})&\leq  BER^{req}\\
     &BER^{Y}(\boldsymbol{V_{bias}^{Y}}, \boldsymbol{G_{amp}^{Y}})&\leq  BER^{req}\\
     & CCT( \boldsymbol{V_{bias}^{R,G,B,Y}})&\leq  CCT^{req}\\
     &\boldsymbol{V_{bias}^{R,G,B,Y}} &\in \{V_{th}^{R,G,B,Y},V_{max}^{R,G,B,Y}\}\\
     &\boldsymbol{G_{amp}^{R,G,B,Y}} &\in \{G_{th}^{R,G,B,Y}, G_{max}^{R,G,B,Y}\}\\ 
\label{c3:opt prob1}
\end{array}
\end{aligned}
\end{equation}

where,  the total illuminance at the user's position ($\Phi_{Rx}^{TL}$) is given by:
\begin{equation}
\begin{aligned}
\Phi_{Rx}^{TL}( \boldsymbol{V_{bias}^{R,G,B,Y}})=&\Phi_{Rx}^{R}(\boldsymbol{V_{bias}^{R}})+\Phi_{Rx}^{G}(\boldsymbol{V_{bias}^{G}})\\
&+\Phi_{Rx}^{B}(\boldsymbol{V_{bias}^{B}})+\Phi_{Rx}^{Y}(\boldsymbol{V_{bias}^{Y}})\\
\label{c3:con phi}
  \end{aligned}
\end{equation}

\begin{equation}
\begin{aligned}
    &BER(\boldsymbol{V_{bias}}, \boldsymbol{G_{amp}})=\frac{4}{\log_{2}(M)}(1-\frac{1}{\sqrt{M}})\\
    &\cdot Q\left(\sqrt{\frac{3\log_{2}(M)}{M-1}\text{SNR}(\boldsymbol{V_{bias}}, \boldsymbol{G_{amp}})}\right)
    \end{aligned}
\end{equation}
\begin{equation}
   Q(x)=\frac{1}{2}\text{erfc}\left(\frac{x}{\sqrt{2}}\right)
   \label{c3:eqQ}
\end{equation}

Here, considering the WDM VLC system using different orders of QAM signal, the BER is evaluated by the order $M$ with the Q-function \cite{c3:berref}. The Q-function represents the tail distribution of the standard normal distribution \cite{c3:Qfunction}, as shown in Eq. \eqref{c3:eqQ}. The expansion of all the constraint equations are derived in the Appendix.

For convenient solving the optimization problem, define the functions $f=-\mathbb{EE^{TL}}$, $g_{1}=\Phi_{Rx}^{req}-\Phi_{Rx}^{TL}$, $g_{2}$ through $g_{5}$  linked to $BER^{R,G,B,Y}-BER^{req}$, and $g_{6}=CCT-CCT^{req}$. The optimization variable and the corresponding boundary set are defined as $X$ and $X_{set}$. The previous expression (Eq. \eqref{c3:opt prob1}) is reformulated into a standard form given by:





\begin{equation}
\begin{aligned} 
&\min_{\boldsymbol{X} \in \boldsymbol{X_{set}}} \quad  f(\boldsymbol{X})\\
&\begin{array}{r@{\quad}r@{}l@{\quad}l}
s.t. & g_{i}(\boldsymbol{X})\leq 0 , i=1,2,...,6\\
\label{c3:stan opt}
\end{array}
\end{aligned}
\end{equation}

\begin{equation}
    G=[\nabla g_{1}(\boldsymbol{X}_{k}),\nabla g_{2}(\boldsymbol{X}_{k}),...,\nabla g_{6}(\boldsymbol{X}_{k})]^{T}
\end{equation}

\begin{equation}
    h=[- g_{1}(\boldsymbol{X}_{k}),- g_{2}(\boldsymbol{X}_{k}),...,- g_{6}(\boldsymbol{X}_{k})]^{T}
\end{equation}

The above nonlinear optimization problem with nonlinear constraints in Eq. \eqref{c3:stan opt} is tentatively solved by the Sequential Quadratic Programming (SQP) algorithm \cite{sqp} to validate our proposed optimization method. The advanced algorithms will be discussed and implemented in our future work. The processing step is structured in Algorithm. \ref{c3:alg:sqp}. 

\begin{algorithm}
\caption{Sequential Quadratic Programming (SQP)}
\begin{algorithmic}[1]
\STATE {\textbf{Initialization:}} Choose an initial feasible point \( \boldsymbol{X}^{(0)} \), the tolerance \(\epsilon\). Set \( k=0 \). Calculate \( \boldsymbol{H}^{(0)} \).
\WHILE{not converged}
\STATE {\textbf{Linearize Constraints and Objective:}} Approximate \( g(\boldsymbol{X})_{i} \) and \( f(\boldsymbol{X}) \)  by  Taylor expansion around \( \boldsymbol{X}^{(k)} \).
\STATE {\textbf{Solve Quadratic Subproblem:}}
\begin{align*}
    \text{minimize}_{ \boldsymbol{p}^{(k)}} \quad & \frac{1}{2}  (\boldsymbol{p}^{(k)})^T \boldsymbol{H} \boldsymbol{p}^{(k)} + \boldsymbol{c}^T \boldsymbol{p}^{(k)} \nonumber \\
    \text{subject to} \quad & \boldsymbol{G}  \boldsymbol{p}^{(k)} \leq \boldsymbol{h} \nonumber
    \end{align*}
    
\STATE {\textbf{Results:} \( \boldsymbol{p}^{(k)} \) and \( \boldsymbol{\mu}^{(k)} \)}
\STATE {\textbf{Line Search:}} Determine \( \alpha^{(k)} \) via line search.
\STATE {\textbf{Update:}} Set \(  \boldsymbol{X}^{(k+1)} =  \boldsymbol{X}^{(k)} + \alpha^{(k)} \boldsymbol{p}^{(k)} \).
\STATE {\textbf{Hessian Update:}} Update \( \boldsymbol{H}^{(k+1)} \).
\STATE {\textbf{Check Convergence:}} If \( \| \nabla f( \boldsymbol{X}^{(k+1)}) \| < \epsilon \), terminate.
\ENDWHILE
\end{algorithmic}
\label{c3:alg:sqp}
\end{algorithm}



where  matrix $\boldsymbol{c}$ is defined as $\boldsymbol{c}=\nabla f(\boldsymbol{X}_{k})$.  $\boldsymbol{p}^{(k)}$ is the optimal step defined as \(\boldsymbol{p}^{(k)}=\boldsymbol{X}-\boldsymbol{X}_k\). 
$\boldsymbol{\mu}$ is the matrix of the Lagrange Multipliers for the six constraints ($g_{1}$ to $g_{6}$).

\(\boldsymbol{H}\) is an approximation to the Hessian matrix of the Lagrangian. The Broyden-Fletcher-Goldfarb-Shanno (BFGS) method \cite{c3:bfgs} is used to update the Hessian approximation $\boldsymbol{H}_{k+1}$. Here, $\boldsymbol{s}_k = \boldsymbol{x}_{k+1} - \boldsymbol{x}_k$ and $\boldsymbol{y}_k = \nabla f(\boldsymbol{x}_{k+1}) - \nabla f(\boldsymbol{x}_k)$. Upon solving each Quadratic Programming (QP) subproblem by the KKT conditions, $ \boldsymbol{p}^{(k)}$ is obtained, which is then used to update the current solution to the next iteration as \(\boldsymbol{X}_{k+1} = \boldsymbol{X}_k + \alpha_k \boldsymbol{p}^{(k)}\). Where \(\alpha_k\) is a step size determined via a line search.


\section{Validation}

This section illustrates the validation process in which the proposed energy efficiency optimization method is evaluated based on an experimental-based simulation.  Initially, the electroluminescent characteristics and signal power loss of the RGBY-LED model are calibrated by experimental measurements. Subsequently, an indoor broadcasting scenario with different use-cases are simulated to assess the performance and generality of the proposed optimization method.

\subsection{RGBY-LED Model Calibration}

A integrated RGBY-LED \cite{c3:jiangfengyi} serves as the sample LED which consists of four monochromatic sub-LEDs, each with its dedicated control pin.  Employing it into a VLC transmission link testbed, the electroluminescent properties of each individual sub-LEDs and the signal power loss across each color channel are measured and fitted with the theoretical derivation.

\subsubsection{RGBY-LED Electroluminescence}

\begin{figure}[htbp]
\centering  
\includegraphics[scale=0.260]{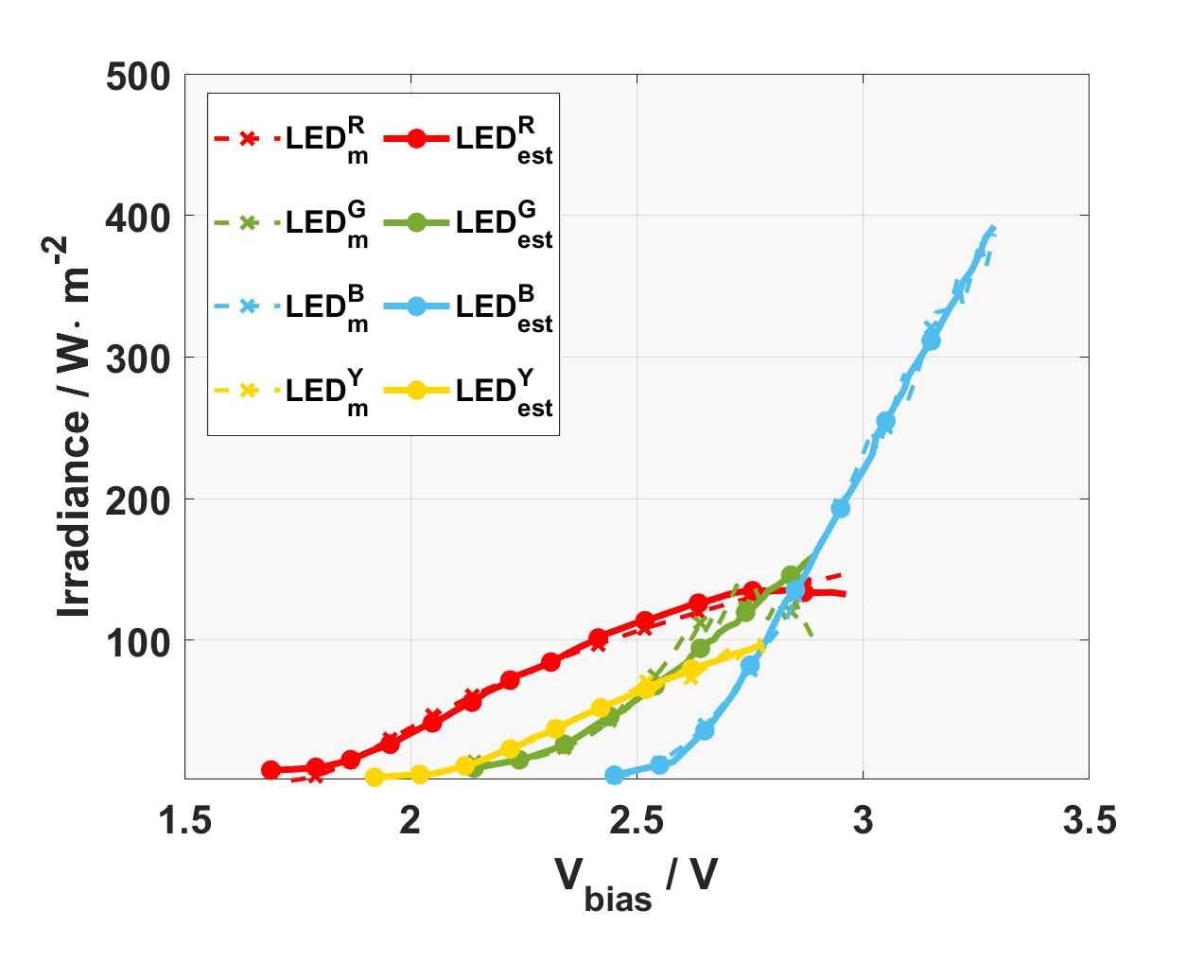}
\caption{Electroluminescent characteristics of the RGBY-LED.}
\label{c3:fig:RGBY EP}
\end{figure}

Each sub-LED is measured separately with the applied bias voltages and output irradiance ($W\cdot m^{-2}$) by an accurate power source in conjunction with a photometer. The measured data are fitted with the theoretical LED model, shown in Fig. \ref{c3:fig:RGBY EP}.

In the figure, the subscript $m$ and $est$ denote the measured and fitted electroluminescence. The fitting results are consistent with the measurement and show the significantly different electroluminescence between each sub-LED. It emphasizes that the unique physical properties of each LED should neither be simplified nor overlooked. A physics-based LED model is significant for any upper-layer analysis of VLC system.

\subsubsection{Signal Power Loss over Channels}

\begin{figure}[htbp]
\centering  
\includegraphics[scale=0.5]{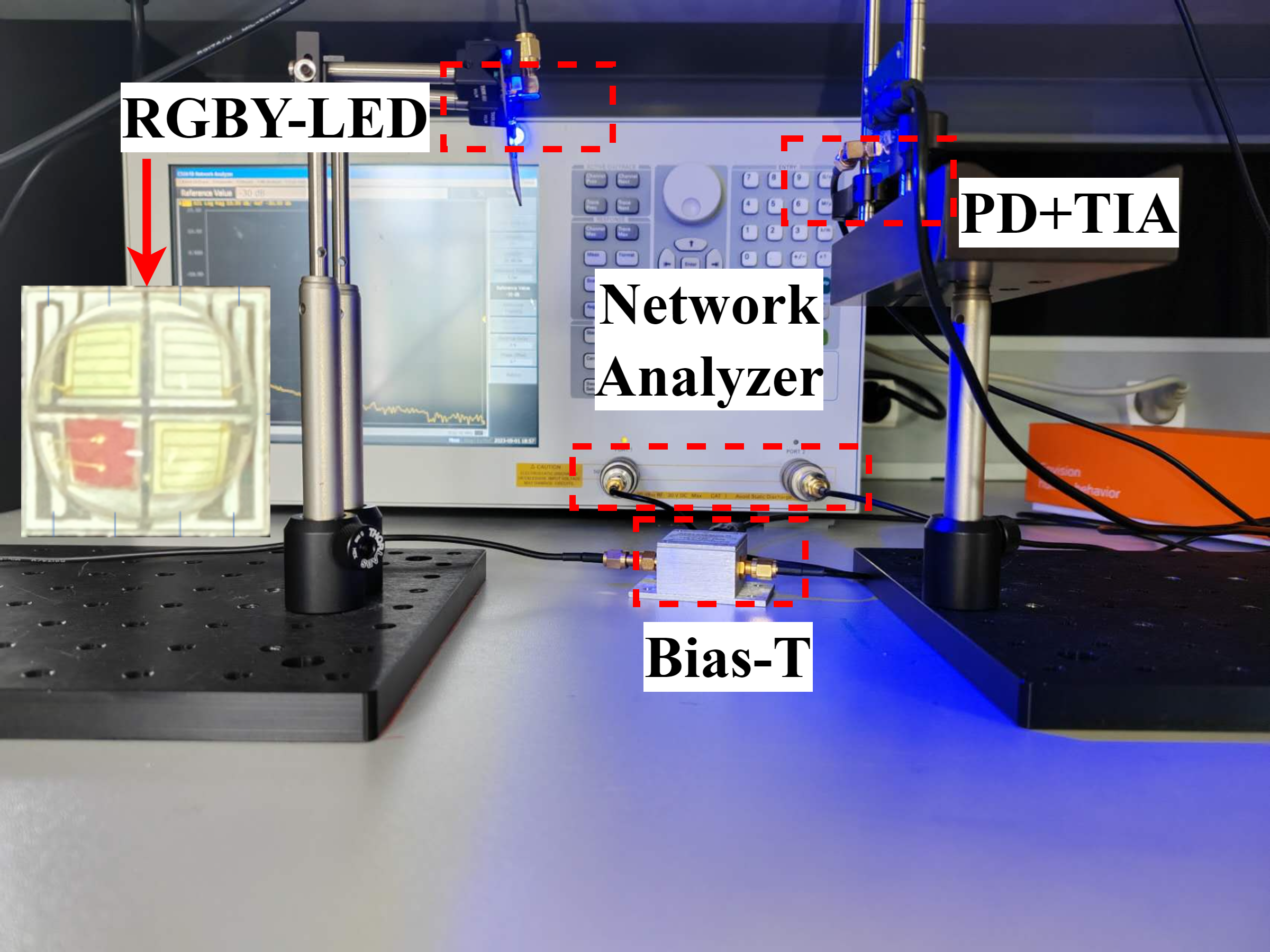}
\caption{VLC transmission link testbed.}
\label{c3:figure: testbed}
\end{figure}

In the VLC transmission testbed, illustrated in Fig. \ref{c3:figure: testbed}, each color channel is separately evaluated for signal power loss using a Network Analyzer. The analyzer generates a frequency-swept signal ranging from 300 KHz to 50 MHz, which is then added on varying bias voltages by a Bias-T. This signal is transmitted over the LED, through the optical wireless channel, and finally received by an optical receiver, consisting of a PD and TIA, to calculate the signal power loss. The distance between the LED and the receiver is fixed on 10 $cm$.

\begin{figure}[htbp]
\centering  
\includegraphics[scale=0.35]{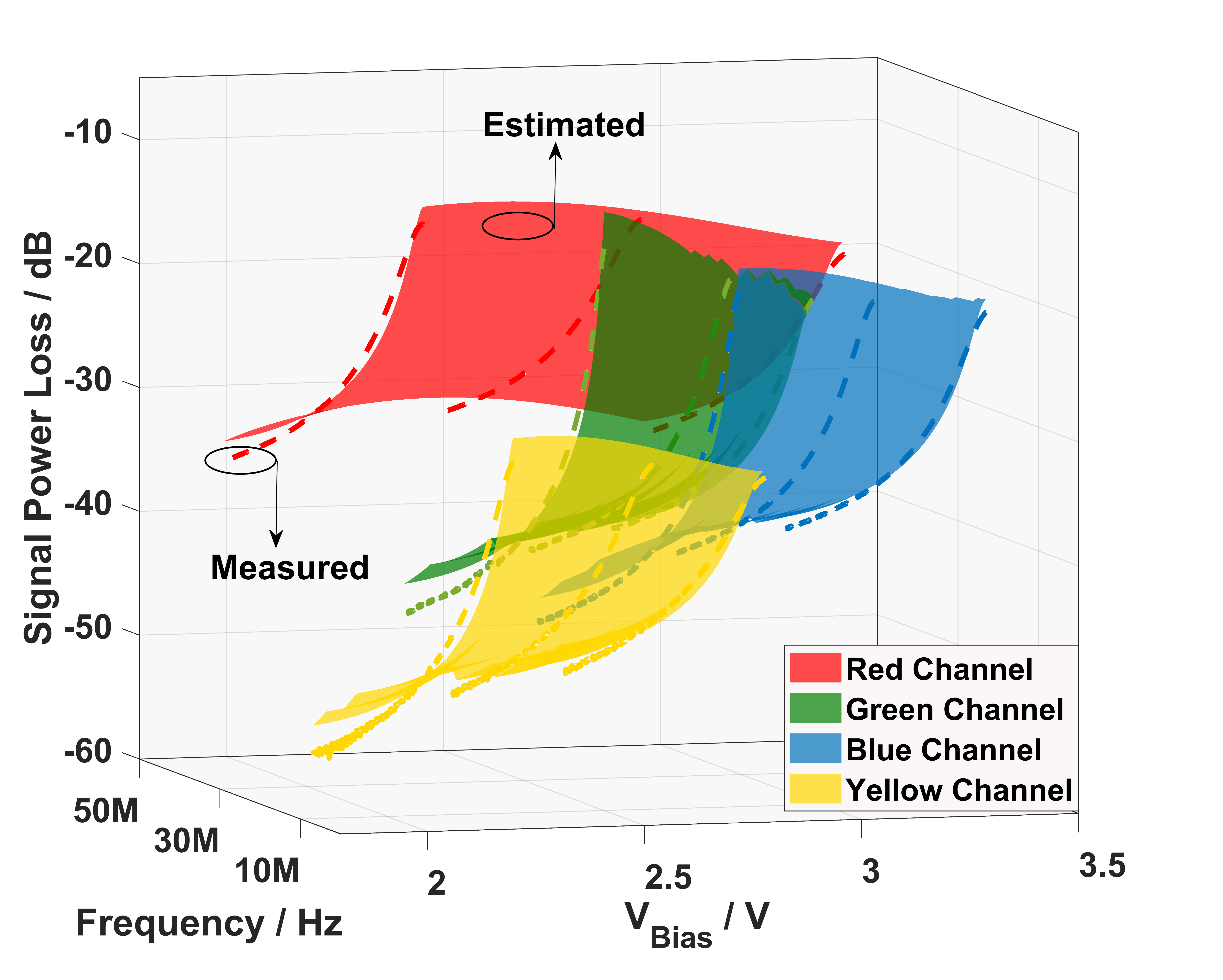}
\caption{Signal power loss of four-color channels.}
\label{c3:figure: Extraction}
\end{figure}

Fig. \ref{c3:figure: Extraction} illustrates the measured and model estimated signal power loss, under varying signal frequency and applied bias voltage, across the four-color channels.  The comparison reveals that the proposed WDM VLC system model accurately characterizes the signal transmission properties for each color channel.  Minor discrepancies between the measured and estimated signal power losses are attributed to the tolerances of the measuring equipment.

The surface trends in the figure validate two key inferences of the proposed system model: (a) Each channel demonstrates a nonlinear frequency attenuation which arises from the unique physical properties of each monochromatic LED. (b) Signal power loss escalates with increasing bias voltage, owing to the interplay of the LED between providing different illumination intensity and transmitting signals. 
The proposed RGBY-LED model accurately characterizes this interplay of each channel, supporting accurate and practical energy efficiency optimization.

\subsection{Energy Efficiency Optimization Simulation}

\begin{figure}[htbp!]
\centering
\includegraphics[scale=0.65]{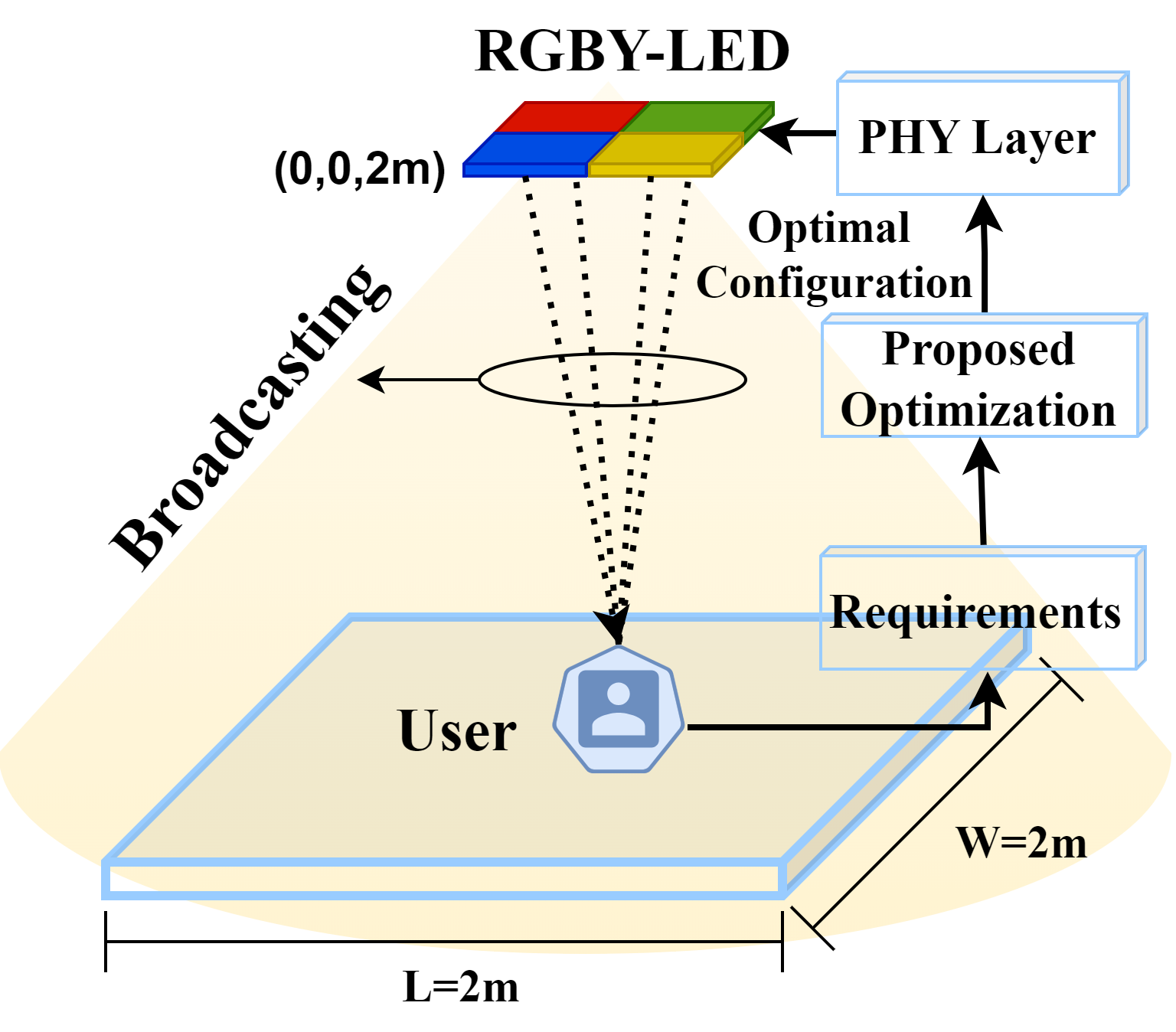}
\caption{Schematic of the indoor WDM VLC broadcasting system. }
\label{c3:fig:setup}
\end{figure}

 The calibrated RGBY-LED model is programmed into the simulation in which an indoor WDM VLC broadcasting system is established, shown in Fig. \ref{c3:fig:setup}. In the simulation, the parameters calibrated from measurements are marked with symbol $^{*}$, listed in the Tab. \ref{c3:tab:simupara}, and the rest are generated following typical values \cite{simu}.

\begin{table}[!ht]
    \centering
    \renewcommand\arraystretch{1.2}
     \setlength{\tabcolsep}{3pt}
    \caption{Parameters of the simulation setup}
    \begin{tabular}{|ll|}
    \hline
        \textbf{Parameters ($^{*}$: Calibrated from measurements)} & \textbf{Values} \\ \hline
        Simulation area & $2m$ $\times$ $2m$ $\times$ $2m$ \\ \hline
        \multirow{4}{*}{$^{*}$Bias voltage of the LED   ($V_{bias}^{R,G,B,Y}$)}& $V_{bias}^{R} \in [1.99,2.96] V$ \\ 
        ~ &$V_{bias}^{G} \in [2.41,2.89] V$ \\    
        ~ & $V_{bias}^{B} \in [2.72,3.29] V$ \\ 
        ~ & $V_{bias}^{Y} \in [2.20,2.78] V$\\ 
        \hline
         \makecell[l]{$^{*}$Central wavelength of the LED's \\ output light ($\lambda_{0}^{R,G,B,Y}$)}& (629,525,460,556) $nm$\\
        \hline 
        \makecell[l]{$^{*}$Optical spectrum bandwidth \\of the LED ($\Delta \lambda^{R,G,B,Y}$)} & (60,65,70,80) $nm$\\
        \hline
        Modulation orders of QAM (M) & 16/32/64\\
        \hline
        Bandwidth of OFDM signals & 3.84 $MHz$\\
        \hline
        Number of subcarriers and OFDM symbols & $256 \times 100$ \\
        \hline
        Initial signal average power ($P_{0}$) & 2 $dBm$\\
        \hline
         The noise floor of the system ($P_{noise}$) & -120 $dBm$\\
        \hline
        Range of amplification gain ($G_{amp}^{R,G,B,Y}$) & [0,20] $dB$\\
        \hline
         Central wavelength of the optical filter ($\hat{\lambda}_{c}$) & (629,525,460,556) $nm$\\
        \hline
         Bandwidth of the optical filter & ideal \\
        \hline
        Coefficient of the optical filter loss ($\Gamma_{p}$)& 0.8\\
        \hline
        Gain of the optical concentrator ($G_{opt}$)& 20 $dB$\\
        \hline
        Lambert radiation coefficient ($m$)& 1.5\\
        \hline
        $^{*}$Effective area of the receiver ($A_{rec}$)& $7.6\times10^{-6} m^{2}$\\
        \hline
        $^{*}$FOV of the LED & $60^\circ$\\
        \hline
        
    \end{tabular}
    
    \label{c3:tab:simupara}
\end{table}

\subsubsection{Simulation Setup}

The simulation background is defined as a user receiving the broadcasting OFDM signals from an RGBY-LED-served WDM VLC system. The broadcasting coverage is a $2m\times2m$ area, and the RGBY-LED is suspended from a $2m$ height ceiling. The proposed optimization method solves the optimal configuration of the system under the input constraints. Then, the solved configuration is returned to the WDM VLC system model to evaluate the system's energy efficiency ($EE^{TL}$). 

 \begin{figure}[htbp!]
\centering
\includegraphics[scale=0.4]{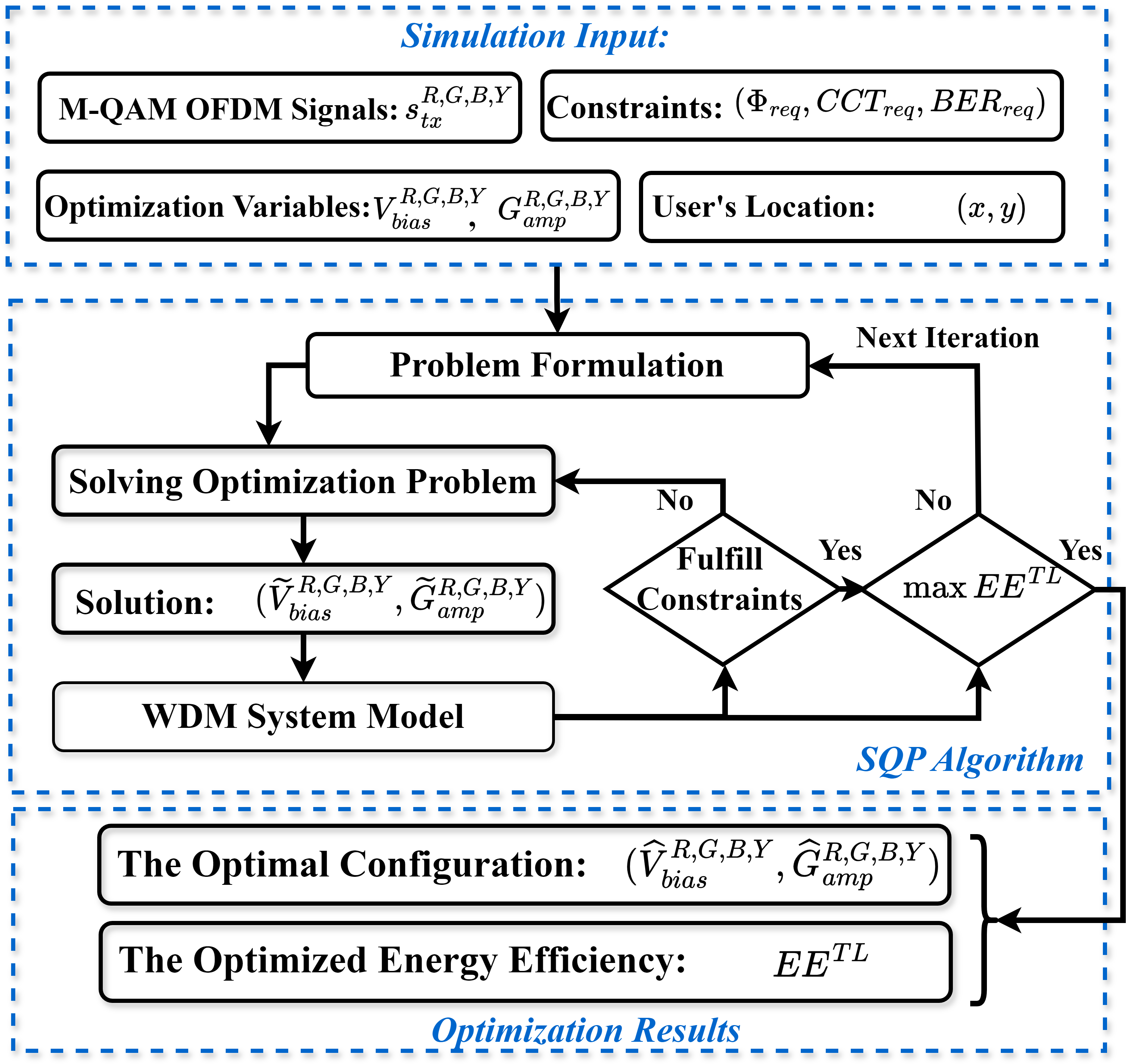}
\caption{Schematic diagram of the simulation process}
\label{c3:fig:process}
\end{figure}

Fig. \ref{c3:fig:process} details the process of the simulation. The variables are generated and bounded following the values in Tab. \ref{c3:tab:simupara}. The user's location is initiated as a uniform distribution within the $2m\times2m$ broadcasting coverage at a $20 cm$ step width. The user's requirements of illuminance ($\Phi_{req}$), illumination color ($CCT_{req}$), and threshold BER of each channel ($BER_{req}$) are quantified into constraints. These quantities form the simulation input. 

Consequently, the formulated optimization problem is solved by the SQP algorithm.  During each iteration, the solved configuration ($\widetilde{V}_{bias}^{R,G,B,Y},\widetilde{G}_{amp}^{R,G,B,Y}$) is assessed through the WDM VLC system model to confirm its adherence to the constraints. Iterations continue until a configuration maximizing the system's energy efficiency is identified, at which point the optimal configuration($\widehat{V}_{bias}^{R,G,B,Y},\widehat{G}_{amp}^{R,G,B,Y}$), and optimized energy efficiency ($EE^{TL}$) are returned.

\begin{figure*}[htbp]
\centering
\subfigure[]{
\includegraphics[scale=0.3]{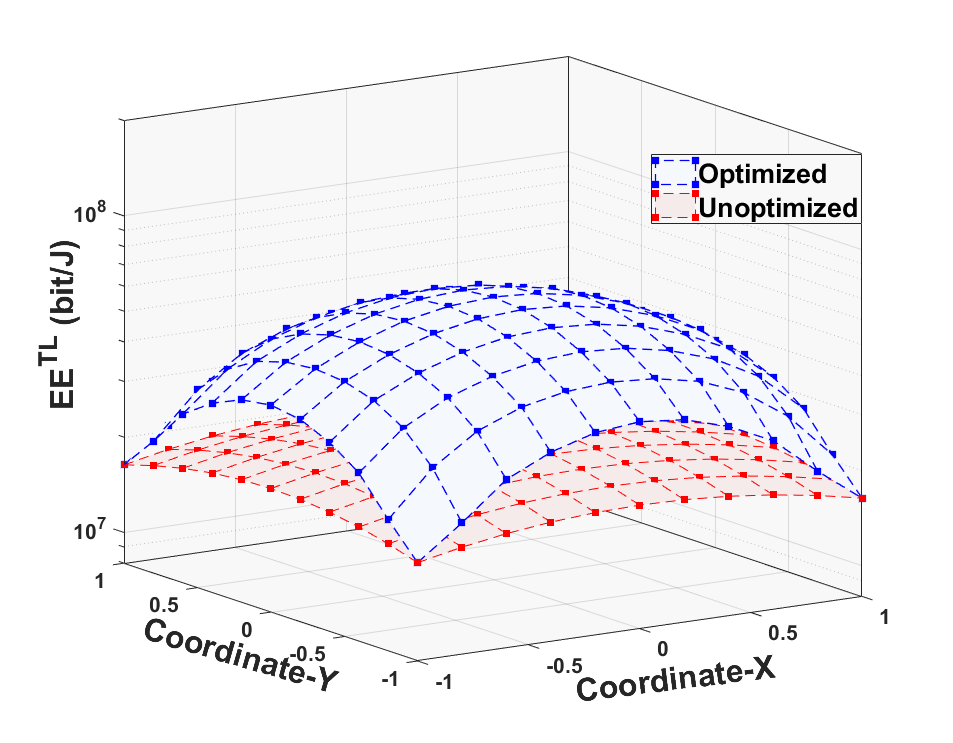}}
\quad
\subfigure[]{
\includegraphics[scale=0.3]{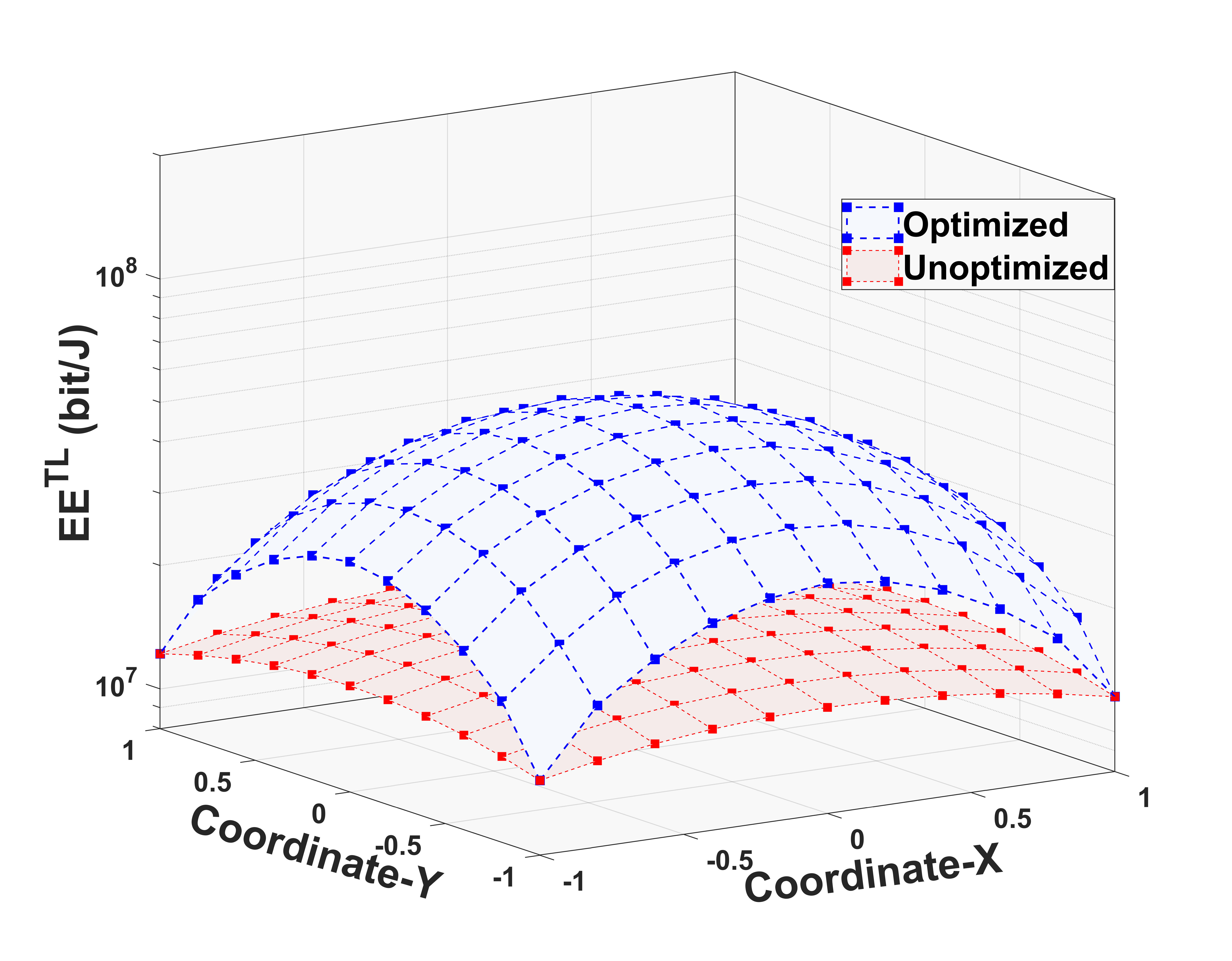}}
\caption{Optimized energy efficiency when the user is at different locations: (a) Transmitted 16-QAM-OFDM signal, and (b) Transmitted 32-QAM-OFDM signal.} 
\label{c3:fig:16 32 mobile}
\end{figure*}

\subsubsection{Optimization Performance}
To validate the performance of the proposed method, each point of the simulated area is traversed to implement a whole process of the optimization and compare the results with the control group, labeled "Optimized" and "Unoptimized" in the legend. The control group behaves as a conventional optimization, in which the system maintains a fixed configuration with minimal power consumption under the constraints. The constraints for illuminance and CCT are constantly set as $\Phi_{Rx}^{req} \geq 200 lx$, and $CCT \leq 5000K$, satisfying the indoor illumination standards for an office environment \cite{c3:cie}. The BER is required lower than $BER^{R,G,B,Y} \leq 10^{-5}$ for each channel, respecting the wireless communication requirement \cite{c3:wdmrgby}.  Fig. \ref{c3:fig:16 32 mobile} displays the optimization results of system transmitting 16-QAM and 32-QAM OFDM signals.

In the Fig. \ref{c3:fig:16 32 mobile} (a), the proposed optimization method achieves the highest energy efficiency of $6.25\times10^{7}$ (bit/J) at the location (0,0), whereas the control group reaches  $2.37\times10^{7}$ (bit/J). At the boundary, both methods yield an energy efficiency of $1.63\times10^{7}$ (bit/J). The average values of each user's location for the proposed method and control group are $4.13\times10^{7}$ (bit/J) and $2.05\times10^{7}$ (bit/J), respectively. In 32-QAM transmission, shown in Fig. \ref{c3:fig:16 32 mobile} (b), the proposed method and the control group achieve peak energy efficiencies of $5.46\times10^{7}$ (bit/J) and $1.74\times10^{7}$ (bit/J) with the average values of each location $3.44 \times10^{7}$ (bit/J) and $1.51 \times10^{7}$ (bit/J), respectively.

 The proposed method calculates the most cost-effective signal power subject to the given constraints, thereby yielding double energy efficiency compared to the conventional approach. Furthermore, the average energy efficiency enhancement of the proposed method increases from 101\% to 127\% with the modulation format from 16-QAM to 32-QAM. Because, under equivalent conditions, a higher data rate demands increased energy consumption, whereas the optimization effect becomes more pronounced.

\subsubsection{Generality Validation}
As the performance of the proposed optimization method is evaluated under a fixed constraints condition, its generality applying to different communication and illumination requirements are validated in this section. Initially, it is assumed that the broadcasting system serves for diverse users, where differentiation between user data is achieved through frequency division multiplexing. Consequently, the proposed optimization method must be adaptable to scenarios wherein the system transmits signals across varying central frequencies. Subsequently, in consideration of the varied illumination demands of users, the proposed method is required to be applicable to scenarios wherein the system provides diverse intensities and visual colors of illumination. 
\subsubsection*{\textbf{Case(1) Different signal frequencies.} }
\begin{figure}[htbp!]
\centering
\includegraphics[scale=0.34]{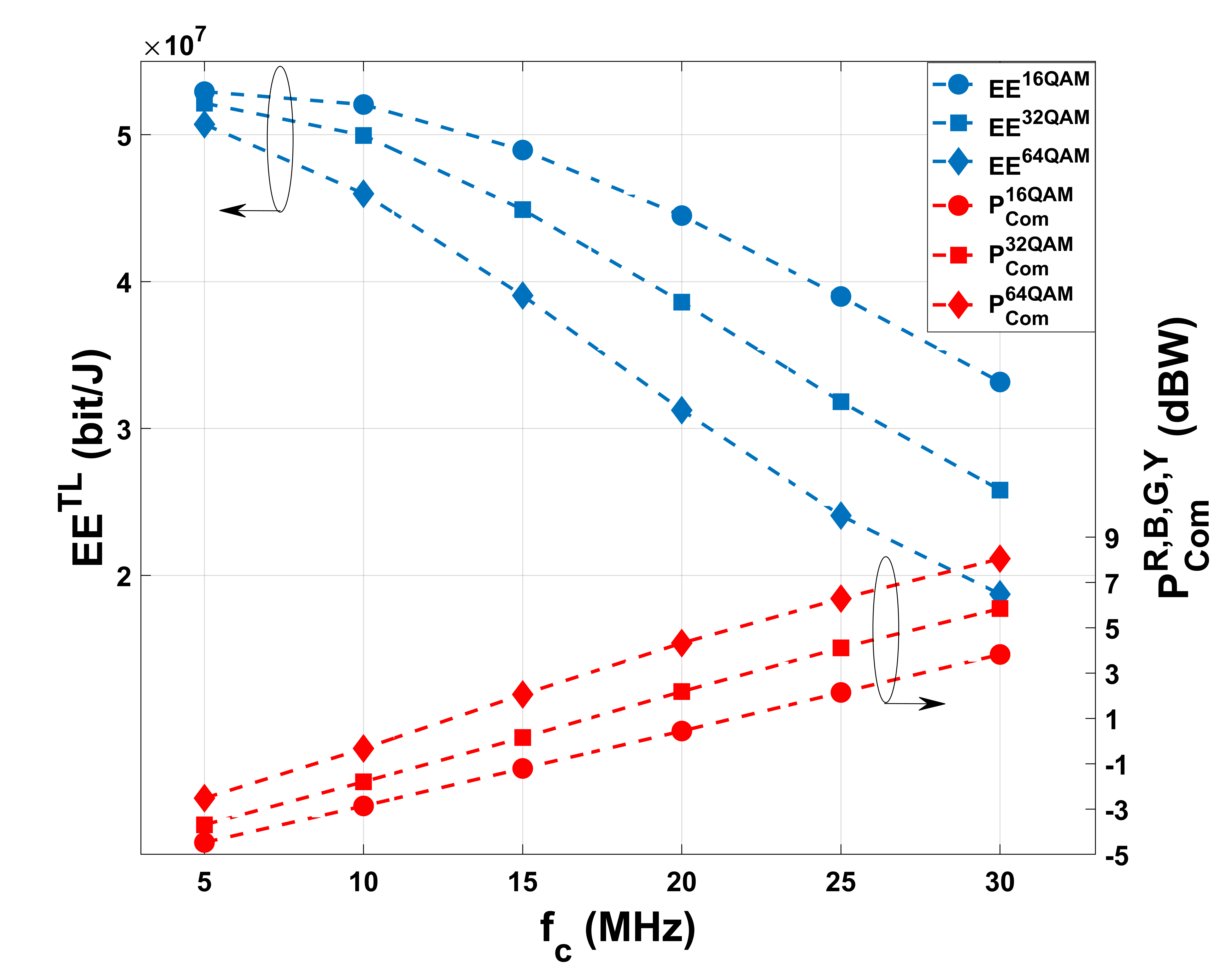}
\caption{Optimized energy efficiency at different signal's central frequency.}
\label{c3:fig:diff fc}
\end{figure}

 The user is fixed at the coordinate $(0,0)$, and the illumination constraints are set at $\Phi_{Rx}^{req} \geq 400 lx$, and $CCT \leq 5000K$, simulating the illumination requirements of an indoor reading scenario. The BER is set as $BER^{R,G,B,Y} \leq 10^{-5}$. OFDM signals for 16-QAM, 32-QAM, and 64-QAM, with central frequencies ranging from 5MHz to 30MHz, are used in the broadcast service. Fig. \ref{c3:fig:diff fc} illustrates optimized energy efficiency ($EE^{TL}$) and corresponding communication power consumption for each central frequency.

Applying the proposed method, optimized energy efficiency $EE^{TL}$ decreases sequentially from $5\times10^{7}$ (bit/J) to $3.32\times10^{7}$ (bit/J), $2.58\times10^{7}$ (bit/J), and $1.87\times10^{7}$(bit/J) for 16-QAM, 32-QAM, and 64-QAM OFDM signal transmission, respectively. Concurrently, the signal power consumption across four channels escalates from $-4.47$ dBW, $-3.69$ dBW, and $-2.52$ dBW to $3.82$ dBW, $5.84$ dBW, $8.04$ dBW, respectively. 

Two key insights are gleaned from these results. First, they confirm the practical understanding that higher data rates necessitate greater signal power. Second, the results align with the measured response of the WDM VLC system, indicating that signal power loss significantly rises with increasing signal frequency. The system's generality across varied transmitted signal frequencies is attributable to the accurate LED model.

\subsubsection*{\textbf{Case(2) Diverse illumination requirements}}

 \begin{figure*}[htbp!]
\centering
\subfigure[]{
\includegraphics[scale=0.34]{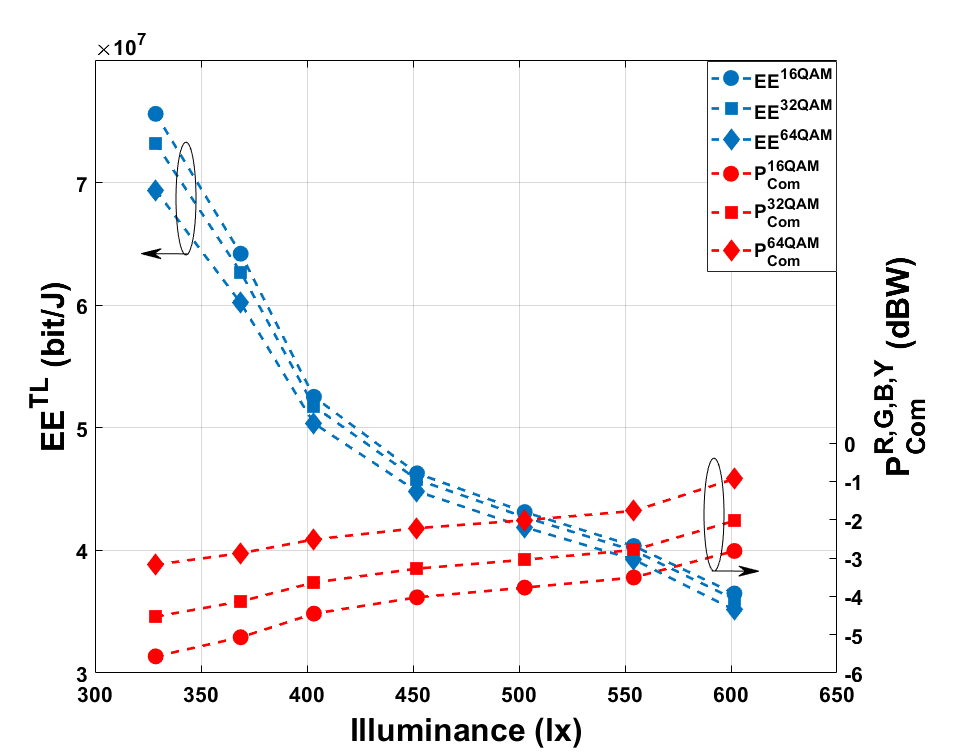}}
\quad
\subfigure[]{
\includegraphics[scale=0.34]{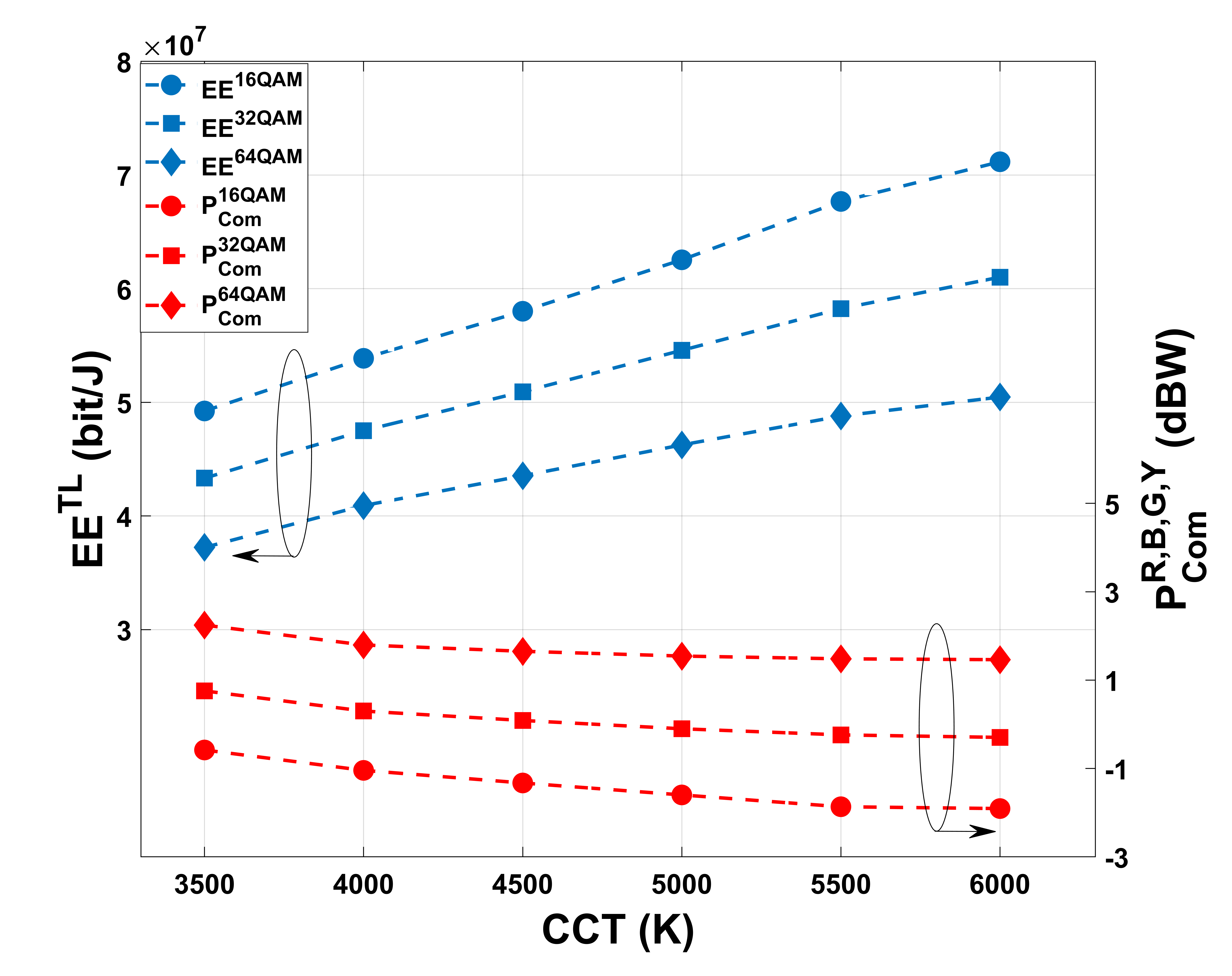}}
\caption{(a) Optimized energy efficiency at different user-required illuminance (b) The optimized energy efficiency at different user-required illumination colors.}
\label{c3:fig:diff ill}
\end{figure*}

To further evaluate the generality of the proposed method on diverse illumination requirements. the user is fixed on the location $(0,0)$,  and the BER is required as $BER^{R,G,B,Y} \leq 10^{-5}$. The constraints of user-required illuminance ($\Phi^{req}$) and illumination color ($CCT^{req}$) vary from  $300lx$ to $600lx$ and $3000K$ to $5000K$, respectively, almost covering the indoor illumination scenarios following the CIE standards \cite{c3:cie}. Fig. \ref{c3:fig:diff ill} (a) and (b) depict the optimized energy efficiency in response to these different illumination requirements, along with the corresponding communication power consumption  ($P_{com}^{R,G,B,Y}$).

In Fig. \ref{c3:fig:diff ill} (a), energy efficiency markedly declines as required illuminance increases. It is noteworthy that the power consumption for communication increases with the required illuminance. This phenomenon is consistent with the measurements shown in Fig. \ref{c3:figure: Extraction} that signal power loss increases with the applied voltage. The system requires more signal power to maintain the response while providing higher illuminance. The mechanism behind this phenomenon lies in the interplay between the illumination and communication functionalities of the LED. 

In Fig. \ref{c3:fig:diff ill} (b), higher $CCT$ values, corresponding to cooler white light, result in improved energy efficiency. It is because higher $CCT$ values permit more blue components contained in the output light of the RGBY-LED. Due to the superior electroluminescence characteristics of the blue sub-LED in the RGBY-LED system, as demonstrated in Fig. \ref{c3:fig:RGBY EP}, when the blue sub-LED predominates in the illumination service, other sub-LEDs require less power, reducing overall illumination power consumption and thereby enhancing system energy efficiency. Meanwhile, it is evident that the communication power consumption is also reduced, which further increases the system's energy efficiency. Because, the reduction of the illumination power consumption requiring lower bias voltages ($V_{bias}^{R,G,Y}$) of the RGBY-LED, resulting in the higher signal response of the system, as Fig. \ref{c3:figure: Extraction} proved. Thus, less signal power is consumed during the data transmission.  This phenomenon also comes from the interplay between the illumination and communication performance of the system. 

In these simulations, the optimization results of the proposed method accordingly vary with the illumination requirements, and the tendency of the variations is consistent with the physical mechanism and the experimental measurements. Therefore, the proposed method can be generalized to optimize the energy efficiency of the WDM VLC system under different scenarios.

\section{Conclusion}
 In this paper, a novel optimization method is proposed for RGBY-LED based WDM VLC systems. A physics-based LED model is integrated with the transmission channel, and WDM receiver models, forming the system model. It characterizes the signal response and illumination performance across different color channels and quantifies the interplay of them. The energy efficiency of the WDM VLC system is defined and formulated into an optimization problem with the constraints following the indoor illumination standard and communication quality. The proposed optimization method, validated on the experimental-based simulation, achieves the double performance of the conventional. Moreover, it exhibits a strong generality to adapt to diverse scenarios, highlighting its potential for practical implementations.

  In traditional communication systems, maximizing data rates and minimizing power consumption are primary objectives. However, VLC brings an additional layer of complexity with the need to meet illumination requirements. The proposed energy efficiency optimization method successfully addresses this critical challenge, demonstrating that the optimal energy efficiency of a VLC system is not merely tied to minimizing power consumption. Instead, it's about striking a balance between communication and illumination demands.


\section*{APENDIX-I}

The parameter functions in the section \ref{sec:A_1} which describes the internal physics of the LED are derived as Eq. \eqref{APP_1} to \eqref{APP_ce}. The definition of the parameters are listed in TAB \ref{definition}. For convenient illustration in section \ref{sec:A_2}, the constraint equations are unexpanded, the original expressions are shown as Eq. \eqref{expan1} to \eqref{expand end}.

\begin{equation}
z_{n}(V_{j})=\frac{ r_{qn}}{n\cdot (r_{n}+r_{b})+r_{qn}}
\label{APP_1}
\end{equation}

\begin{equation}
n^{*}=\frac{m^{*}k_{B}T}{\pi\hbar^{2}L_{q}}
\end{equation}

\begin{equation}
a_{1}(V_{j})=\eta_{ph}\beta_{sp}nR_{ph}z_{n}(V_{j})
\end{equation}

\begin{equation}
a_{2}=R_{ph}C_{ph}
\end{equation}

\begin{equation}
a_{3}(V_{j})=(r_{q}+n\cdot z_{n})\cdot (Z_{s}+R_{s})+r_{q}z_{n}
\end{equation}

\begin{equation}
a_{4}(V_{j})=r_{q}z_{n}C_{e}(Z_{s}+R_{s})
\end{equation}

\begin{equation}
\begin{aligned}
&a_{5}(V_{j})= n\eta_{ph}R_{ph} \beta_{sp}qA_{q}L_{q}\gamma_{2} (n^{*})^{2}\\
\cdot &\ln\Bigg[1+\exp\bigg(\frac{\alpha_{1}V_{j}+\alpha_{2}V_{j}^{2}+\alpha_{3}V_{j}^{3}}{k_{B}T}\bigg)\Bigg]^{2}
\label{APP_5}
\end{aligned}
\end{equation}

\begin{figure*}
{\noindent} \rule[-10pt]{18cm}{0.05em}
\begin{align}
r_{n}(V_{j})=\frac{\eta k_{B}T}{n_{0}q^{2}A_{c}L_{c}\exp{(\frac{qV_{j}}{\eta k_{B}T})}(\gamma_{1}+3\gamma_{3}\cdot n_{0}^{2}\exp^{2}(\frac{qV_{j}}{\eta k_{B}T}))}
\end{align}
\end{figure*}

\begin{figure*}
\begin{align}
r_{b}(V_{j})=&\frac{\tau_{b}}{q\cdot n^{*}A_{b}L_{b}}\Bigg[\frac{\exp(\frac{\alpha_{1}V{j}+\alpha_{2}V_{j}^{2}+\alpha_{3}V_{j}^{3}}{k_{B}T})\cdot (\alpha_{1}+2\alpha_{2}V_{j}+3\alpha_{3}V_{j}^{2})}{k_{B}T[1+\exp(\frac{\alpha_{1}V_{j}+\alpha_{2}V_{j}^{2}+\alpha_{3}V_{j}^{3}}{k_{B}T})]}\cdot \exp(\frac{q(V_{j}-V_{D})}{k_{B}T})\\
&+\frac{q}{k_{B}T}\cdot \ln[1+\exp(\frac{\alpha_{1}V_{j}+\alpha_{2}V_{j}^{2}+\alpha_{3}V_{j}^{3}}{k_{B}T})]\cdot \exp(\frac{q(V_{j}-V_{D})}{k_{B}T})\Bigg]^{-1}
\end{align}
\end{figure*}

\begin{figure*}
\begin{align}
r_{qn}(V_{j})=\frac{k_{B}T[1+\exp(\frac{\alpha_{1}V_{j}+\alpha_{2}V_{j}^{2}+\alpha_{3}V_{j}^{3}}{k_{B}T})]}{qn^{*}A_{q}L_{q}\exp(\frac{\alpha_{1}V_{j}+\alpha_{2}V_{j}^{2}+\alpha_{3}V_{j}^{3}}{k_{B}T})(\alpha_{1}+2\alpha_{2}V_{j}^{2}+3\alpha_{3}V_{j}^{2})} \cdot \frac{1}{\gamma_{1}+3\gamma_{3}[n^{*}\ln(1+\exp(\frac{\alpha_{1}V_{j}+\alpha_{2}V_{j}^{2}+\alpha_{3}V_{j}^{3}}{k_{B}T}))]^{2}}
\end{align}
\end{figure*}

\begin{figure*}
\begin{align}
r_{qr}(V_{j})=\frac{k_{B}T[1+\exp(\frac{\alpha_{1}V_{j}+\alpha_{2}V_{j}^{2}+\alpha_{3}V_{j}^{3}}{k_{B}T})]}{2(n^{*})^{2}{qn^{*}A_{q}L_{q}\exp(\frac{\alpha_{1}V_{j}+\alpha_{2}V_{j}^{2}+\alpha_{3}V_{j}^{3}}{k_{B}T})(\alpha_{1}+2\alpha_{2}V_{j}^{2}+3\alpha_{3}V_{j}^{2})}}\cdot \frac{1}{\ln(1+\exp(\frac{\alpha_{1}V_{j}+\alpha_{2}V_{j}^{2}+\alpha_{3}V_{j}^{3}}{k_{B}T}))}
\end{align}
\end{figure*}

\begin{figure*}
\begin{align}
C_{e}(V_{j})&=\frac{q^{2}n_{0}A_{c}L_{c}}{\eta k_{B}T}\exp(\frac{qV_{j}}{\eta k_{B}T})+n\cdot \frac{qA_{q}L_{q}n^{*}\exp(\frac{\alpha_{1}V_{j}+\alpha_{2}V_{j}^{2}+\alpha_{3}V_{j}^{3}}{k_{B}T})}{k_{B}T(1+\exp(\frac{\alpha_{1}V_{j}+\alpha_{2}V_{j}^{2}+\alpha_{3}V_{j}^{3}}{k_{B}T}))}\\
& \cdot (\alpha_{1}+2\alpha_{2}V_{j}+3\alpha_{3}V_{j}^{2})+A_{eff}\bigg[\frac{q\epsilon_{q}\epsilon_{b}N_{A}N_{D}}{2(\epsilon_{q}N_{D}+\epsilon_{b}N_{A})}\frac{1}{V_{D}-V_{j}}\bigg]^{\frac{1}{2}}
\label{APP_ce}
\end{align}
\end{figure*}

\begin{figure*}
\begin{align}
& \Phi_{Rx}(\boldsymbol{V_{bias}})=638 \cdot \eta_{ph}R_{ph} \cdot n\beta_{sp}qA_{q}L_{q}\gamma_{2} \cdot \frac{(\mu+1)\cdot cos(\psi)^{\mu}cos(\theta)}{2\pi D^2} \cdot\int_{380}^{830}\mathcal{V}(\lambda) \exp\bigg(\frac{-4(\lambda-\lambda_{0})^{2}}{(\Delta \lambda)^{2}}\bigg) d\lambda \notag\\
& \cdot \Bigg[n^{*}\ln\bigg(1+\exp\bigg(\frac{\alpha_{1}\boldsymbol{V_{bias}}-I_{DC}R_{s}+\alpha_{2}(\boldsymbol{V_{bias}}-I_{DC}R_{s})^{2}+\alpha_{3}(\boldsymbol{V_{bias}}-I_{DC}R_{s})^{3}}{k_{B}T}\bigg)\bigg)\Bigg]^{2}
\label{expan1}
\end{align}
\end{figure*}

\begin{figure*}
\begin{align}
  &BER(\boldsymbol{V_{bias}}, \boldsymbol{G_{amp}})=\frac{4}{\log_{2}(M)}(1-\frac{1}{\sqrt{M}})Q\left(\sqrt{\frac{3\log_{2}(M)}{M-1}\text{SNR}}\right)= \frac{4}{\log_{2}(M)}(1-\frac{1}{\sqrt{M}}) Q\Bigg\{\sqrt{\frac{3\log_{2}(M)}{(M-1)(\sigma_{th}^{2}+\sigma_{shot}^{2})}}\notag\\
&\cdot \frac{\boldsymbol{G_{amp}}\cdot G_{opt}\cdot a_{1}(\boldsymbol{V_{bias}}-I_{DC}R_{s}) }{a_{4}(\boldsymbol{V_{bias}}-I_{DC}R_{s})-a_{2}\cdot a_{3}(\boldsymbol{V_{bias}}-I_{DC}R_{s})} \cdot \frac{(\mu+1)A_{rec}\cdot cos(\psi)^{\mu}cos(\theta)}{2\pi D^2} \cdot \int_{\lambda_{min}}^{\lambda_{max}}\Gamma_{p}\mathrm{rect}(\frac{\lambda-\lambda_{c}}{\Delta \lambda}) \cdot \notag\\
&\exp \bigg(\frac{-4(\lambda-\lambda_{0})^{2}}{(\Delta \lambda)^{2}}\bigg)
\cdot \kappa(\lambda)  d\lambda \cdot \int_{-\infty}^{+\infty}  \Bigg[\exp\bigg(\frac{-a_{3}(\boldsymbol{V_{bias}}-I_{DC}R_{s})\cdot t}{a_{4}(\boldsymbol{V_{bias}}-I_{DC}R_{s})}\bigg)-\exp\bigg(\frac{-t}{a_{2}}\bigg)\Bigg] \cdot s_{tx}(t-\tau)d\tau \Bigg \} 
\end{align}
\end{figure*}

\begin{figure*}
\begin{align}
  CCT( \boldsymbol{V_{bias}^{R,G,B,Y}})=437\cdot \hat{n}^3( \boldsymbol{V_{bias}^{R,G,B,Y}}) + 3601 \cdot \hat{n}^2( \boldsymbol{V_{bias}^{R,G,B,Y}}) + 6861\cdot \hat{n}( \boldsymbol{V_{bias}^{R,G,B,Y}}) + 5517
\end{align}
\label{c3:con CCT}
\end{figure*}

\begin{figure*}
\begin{align}
  \hat{n}( \boldsymbol{V_{bias}^{R,G,B,Y}})&=\frac{\int_{380}^{830}\bar{x}(\lambda)P_{L}^{TL}(\lambda,\boldsymbol{V_{bias}^{R,G,B,Y}})d\lambda}{\int_{380}^{830}P_{L}^{TL}(\lambda,\boldsymbol{V_{bias}^{R,G,B,Y}})(\bar{x}(\lambda)+\bar{y}(\lambda)+\bar{z}(\lambda))d\lambda}-0.3320  \notag\\
  &\cdot \bigg( 0.1858-\frac{\int_{380}^{830}\bar{y}(\lambda)P_{L}^{TL}(\lambda,\boldsymbol{V_{bias}^{R,G,B,Y}})d\lambda}{\int_{380}^{830}P_{L}^{TL}(\lambda,\boldsymbol{V_{bias}^{R,G,B,Y}})(\bar{x}(\lambda)+\bar{y}(\lambda)+\bar{z}(\lambda))d\lambda}\bigg)^{-1}
    \label{expand end}
  \end{align}
{\noindent} \rule[-10pt]{18cm}{0.05em}
\end{figure*}

\begin{table}[htbp!]
\caption{Definition of Terms.}
\label{definition}
\centering
\begin{tabular}{|l|l|}
\hline
Terminology& Definition\\
\hline
$A_{c,q,b}$& Transection area of each layer\\
$A_{eff}$& Effective area of barrier\\
$A_{rec}$& Area of photo-receiver\\
$\alpha_{1,2,3}$& Coefficient of the fitting curve\\
$\beta_{sp}$ & Spontaneous emission coefficient\\
$C$ & Capacitance\\
$D$ & Distance between LED and VLC receiver\\
$\gamma_{1}$ & Coefficient of the SRH recombination\\
$\gamma_{2}$ & Coefficient of the radiative recombination\\
$\gamma_{3}$ & Coefficient of the Auger recombination\\
$\epsilon_{q,b}$ & Relative dielectric constant of the layer\\
$\Phi_{\frac{1}{2}}$ & Semi-angle at half-power of LED\\
$\hbar$& Reduced Planck constant\\
$I_{c,q,b}$& Current of each layer\\
$I_{j}$& Injected current of LED\\
$I_{s}$& Reverse bias saturation current \\
$k_{B}$& Boltzmann constant\\
$\kappa$& Sensitivity of APD\\
$L$& Thickness of the layer\\
$n$& number of the quantum wells \\
$\mu$& Order of Lambertian radiation\\
$N_{D}^{+}$ & Ionized donors density\\
$N_{A}^{+}$ & Ionized acceptors density\\
$\eta$ & Ideal factor of diode\\
$\eta_{L}$ & Efficacy of LED\\
$\eta_{ph}$ & Light extraction rate \\
$P$ & Power\\
$q$ & Elementary charge\\
$\theta$& Angle of receiving direction in VLC channel\\
$R$& Constant resistance\\
$r$& Differential resistance\\
$\sigma_{th,short}$& Thermal and short noise power\\
$T$& Temperature\\
$\tau_{b}$ & Effective space transport time\\
$\tau_{ph}$ & Photons lifetime\\
$V_{D}$ & Potential difference of the barrier\\
$V_{j}$ & Junction voltage\\
$\psi$ & Angle of transmission direction in VLC channel\\
$Z$ & Impedance\\
\hline
\end{tabular}
\end{table}


 
%

\newpage
\bibliography{Ref}
\bibliographystyle{IEEEtran}


 





\end{document}